\documentclass[fleqn,usenatbib]{mnras}

\usepackage{newtxtext,newtxmath}
\usepackage[ampersand]{easylist}
\usepackage{ulem}
\usepackage{verbatim}

\usepackage[T1]{fontenc}
\usepackage{ae,aecompl}

\usepackage{graphicx}	
\usepackage{amsmath}	
\usepackage{amssymb}	
\usepackage{color}
\usepackage[colorinlistoftodos]{todonotes}

\newcommand{\zs}{$z$$\sim$}

\newcommand{\U}{$U$}
\newcommand{\newu}{$u$}
\newcommand{\oldu}{$u^*$}

\newcommand{\grizy}{$grizy$}

\newcommand{\sqdeg}{deg$^2$}

\newcommand{\hscs}{\mbox{HSC-SSP}}
\newcommand{\HSCS}{\mbox{HSC-SSP}}

\title[CLAUDS]{The CFHT Large Area $U$-band Deep Survey (CLAUDS)}
\author[M.\ Sawicki et al.]{
\hspace{-2.0mm}
Marcin Sawicki$^{1}$\thanks{E-mail: {\tt marcin.sawicki@smu.ca}}\thanks{Canada Research Chair},
Stephane Arnouts$^{2}$,
Jiasheng Huang$^{3,4}$,
Jean Coupon$^{5}$,
\newauthor
Anneya Golob$^{1}$,
Stephen Gwyn$^{6}$,
Sebastien Foucaud$^{7}$,
Thibaud Moutard$^{1}$,
\newauthor 
Ikuru Iwata$^{1,8,9}$,
Chengze Liu$^{7}$,
Lingjian Chen$^{1}$,
Guillaume Desprez$^{5}$,
\newauthor 
Yuichi Harikane$^{10}$,
Yoshiaki Ono$^{10}$,
Michael A.\ Strauss$^{11}$,
Masayuki Tanaka$^{8,9}$,
\newauthor
Nathalie Thibert$^{1}$, 
Michael Balogh$^{12}$, 
Kevin Bundy$^{13,14}$, 
Scott Chapman$^{1,15}$, 
\newauthor
James E.\ Gunn$^{11}$, 
Bau-Ching Hsieh$^{16}$,
Olivier Ilbert$^{2}$,
Yipeng Jing$^{7}$, 
\newauthor
Olivier LeF\`evre$^{2}$,
Cheng Li$^{7}$, 
Yuichi Matsuda$^{8,9}$,
Satoshi Miyazaki$^{8,9}$,
Tohru Nagao$^{17}$,
\newauthor
Atsushi J.\ Nishizawa$^{18}$,
Masami Ouchi$^{10}$,
Kazuhiro Shimasaku$^{19}$,
John Silverman$^{13}$,
\newauthor
Sylvain de la Torre$^{2}$,
Laurence Tresse$^{2,20}$, 
Wei-Hao Wang$^{16}$,
Chris J.\ Willott$^{6}$, 
\newauthor
Toru Yamada$^{21,22}$, 
Xiaohu Yang$^{7}$,
Howard K.C.\ Yee$^{23}$
\\
\\
Affiliations are listed in Appendix~\ref{sec:affiliations}
\vspace{-5mm}
}

\date{Accepted 2019 September 3. Received 2019 August 8; in original form 2019 June 19}

\pubyear{2019}

\begin{document}
\label{firstpage}
\pagerange{\pageref{firstpage}--\pageref{lastpage}}
\maketitle
\begin{abstract}
The Canada-France-Hawaii Telescope (CFHT) Large Area $U$-band Deep Survey (CLAUDS) uses data taken with the MegaCam mosaic imager on CFHT to produce images of 18.60~deg$^2$ with median seeing of FWHM=0.92\arcsec\ and to a median depth of $U = 27.1$ AB (5$\sigma$ in 2\arcsec\ apertures), with selected areas that total 1.36~deg$^2$ reaching a median depth of $U=27.7$ AB.  These are the deepest $U$-band images assembled to date over this large an area.   These data are located in four fields also imaged to comparably faint levels in $grizy$ and several narrowband filters as part of the Hyper Suprime-Cam (HSC) Subaru Strategic Program (\HSCS). These CFHT and Subaru datasets will remain unmatched in their combination of area and depth until the advent of the Large Synoptic Survey Telescope (LSST). This paper provides an overview of the scientific motivation for CLAUDS and gives details of the observing strategy, observations, data reduction, and data merging with the \HSCS. Three early applications of these deep data are used to illustrate the potential of the dataset: deep \U-band galaxy number counts, \zs3 Lyman break galaxy (LBG) selection, and photometric redshifts improved by adding CLAUDS \U\ to the Subaru HSC $grizy$ photometry.
\end{abstract}



\section{Introduction}\label{sec:introduction}

A major achievement of observational cosmology has been the detailed characterization of  large scale structure (LSS) traced by galaxies in both the local and high-$z$ Universe, and its successful explanation in the  $\Lambda$CDM cosmological framework.  However, within this broad $\Lambda$CDM paradigm, our understanding of the formation of structures on the scales of galaxies contains vast gaps:  we still do not fully understand the link between galaxy properties and their host dark matter halos or the baryonic physics from gaseous infall, through star formation (SF), to SF-regulating feedback processes. We do not yet fully know how the galaxies around us today evolved from the objects we see at high redshifts. 

A powerful approach to tackle such issues, spectacularly demonstrated at low redshift by the Sloan Digital Sky Survey (SDSS; \citealt{yor00}),  is with deep, large-area imaging and spectroscopic surveys. Such surveys make it possible to derive galaxy properties --- such as stellar mass ($M_{\star}$), star formation rate (SFR), and morphology --- for galaxies drawn from large, representative volumes, and then to use statistical tools to link these properties to those of their host DM halos and to understand the role played by local and large-scale environment. 

However, because galaxies evolve over billions of years, truly large-volume surveys (even the $\sim$2~\sqdeg\ COSMOS field (\citealt{Scoville2007}) is affected by cosmic variance ---  see, e.g., \citealt{arc13}) need to be extended beyond the local Universe to span the vast lookback times over which this evolution takes place.  This extension to high redshift is one of the key motivations for the Hyper Suprime-Cam Subaru Strategic Program (\hscs; \citealt{aih17overview}) currently underway on the Subaru telescope using the Hyper Suprime-Cam imager (HSC; \citealt{miyazaki2018}), as well as for its planned follow-up with the Prime Focus Spectrograph (PFS; \citealt{tak14}).

Of particular interest here are the Deep and UltraDeep components of the HSC-SSP, which image the sky in five broadband ($grizy$) and four narrowband (NB) filters ($\lambda$=387, 816, 921, 1010~nm) to unprecedented combinations of depth and area:  when completed, the Deep component will reach $i_{lim} \sim 27.1$ AB over $\sim$26 \sqdeg\ (5$\sigma$ limits in 2\arcsec\ apertures) and UltraDeep will reach $i_{lim} \sim 27.7$ AB  over 3.5 \sqdeg\ \citep[see][]{Aihara2019DR2}.  The four Deep fields (E-COSMOS, XMM-LSS, ELAIS-N1, DEEP2-3), along with the two UltraDeep fields (COSMOS and SXDS, which are embedded within two of the Deep fields, namely E-COSMOS and XMM-LSS, respectively) are in well-studied areas of the sky rich in ancillary data, including extensive spectroscopy and IR and X-ray imaging.   Together, these data are ideal for galaxy evolution studies as they provide an unprecedented combination of depth and area that is key for assembling large samples spanning a range of properties in a variety of environments, dominating over cosmic variance, beating down statistical noise, and finding rare objects. However, these areas lack $U$-band data of comparable depth and areal coverage. 

\begin{figure}
\begin{center}
\includegraphics[width=0.52\textwidth]{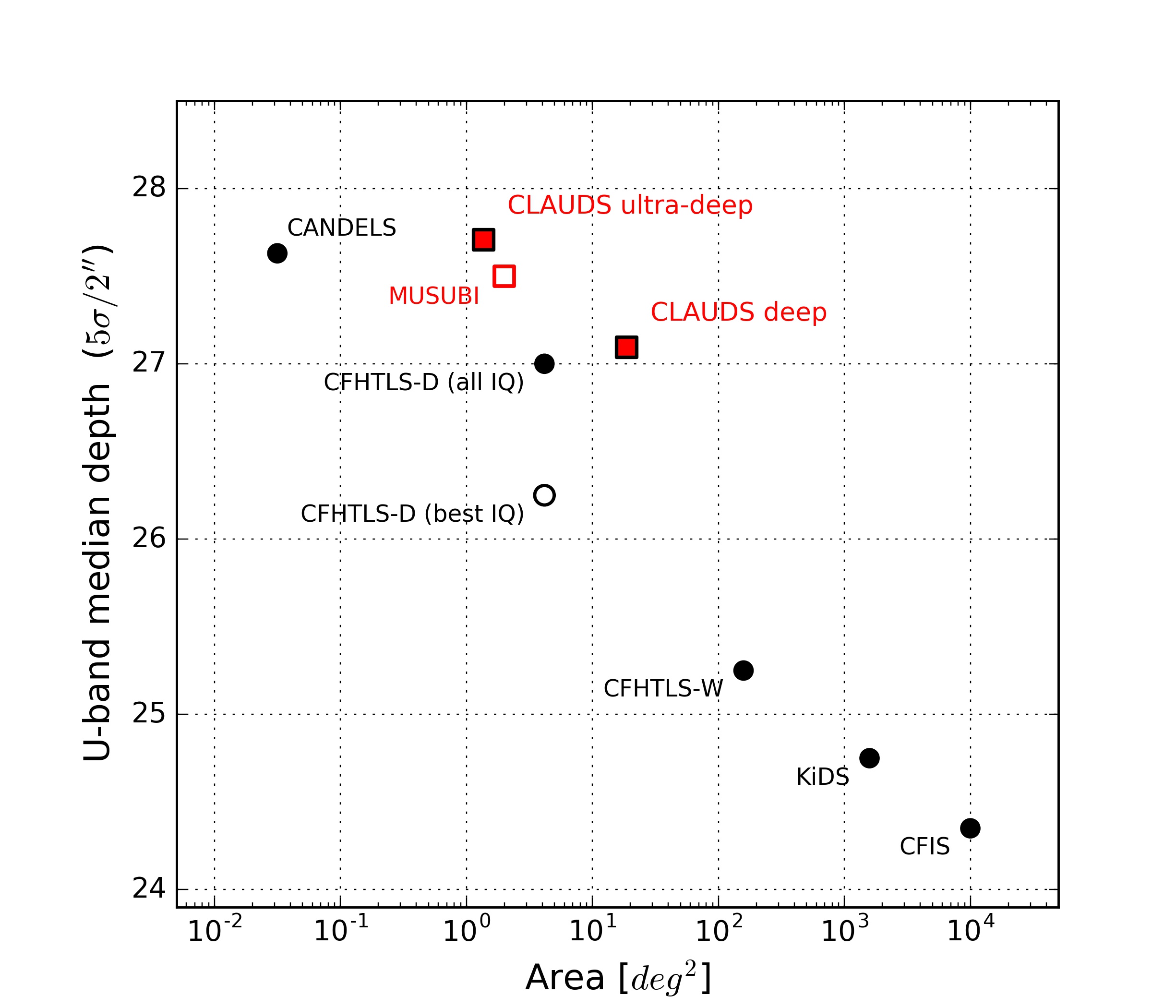}
\end{center}
\caption[]{ 
CLAUDS median depths are shown in the context of recent and ongoing $U$-band surveys:  Cosmic Assembly Near-IR Deep Extragalactic Legacy Survey (CANDELS; \citealt{grogin2011}); the Wide and Deep components of the CFHT Legacy Survey (CFHTLS; \citealt{hud12}) -- CFHTLS-W and CFHTLS-D, respectively; the Kilo Degree Survey (KiDS; \citealt{deJong2017}); and the Canada-France Imaging Survey (CFIS, \citealt{Ibata2017}).   The red open square is for the MegaCam Ultra-deep Survey with U-Band Imaging (MUSUBI; W.-H. Wang et al., in preparation) whose data are incorporated into the CLAUDS stacks. Note that CFTHLS-D and MUSUBI contain CFHT \oldu\ data in the COSMOS and SXDS fields that we incorporate into our CLAUDS stacks. The CLAUDS ultra-deep area point includes both \newu\ and \oldu\ stacks in the central COSMOS field in addition to the \oldu\ data in the SXDS field that forms part of our XMM-LSS field. 
}
\label{ubandSurveys.fig}
\end{figure}

While we can learn much from data at longer wavelengths, $U$-band observations are critical for several areas of research, including for studies at $z \la 0.7$ and $z\sim2-3$, as we discuss in this paper.  A number of deep and wide $U$-band imaging surveys have been undertaken in the past (see Fig.~\ref{ubandSurveys.fig}), including some that overlap parts of the HSC-SSP Deep and UltraDeep fields, but none of them have the combination of depth and area needed to match the HSC-SSP Deep and UltraDeep $grizy$+NB observations. 
\U-band data of depth comparable to the \hscs\ images are thus essential, and with this in mind we  carried out the CFHT Large Area $U$-band Deep Survey (CLAUDS). This survey was enabled by the MegaCam imager \citep{boul03} which, in contrast to HSC and most other imagers on large-aperture telescopes, is unique in its combination of areal coverage ($\sim$$1 \deg^2$) and \U-band sensitivity. The stacked CLAUDS images that we produced reach a median depth of $U = 27.1$ AB (5$\sigma$ in 2\arcsec apertures)  over 18.60 $\deg^2$ in the HSC-SSP Deep layer, with selected areas within the UltraDeep regions that total 1.36~deg$^2$ reaching a median depth of $U=27.7$ AB. 

As a stand-alone survey CLAUDS is thus unmatched in the space of \U-band surveys (see Fig.~\ref{ubandSurveys.fig}). Combined, CLAUDS+\HSCS\ will be unsurpassed in their combination of depth, area and wavelength coverage until the advent of the Large Synoptic Survey Telescope (LSST; \citealt{Ivezic2019}, \citealt{lsst2009}). Together, these data will allow an unprecedented exploration of cosmic evolution at $0 \la z \la 3$.   Even in the era of highly-multiplexed spectroscopy, photometric redshifts (photo-$z$) are essential for large and deep galaxy samples.  At intermediate redshifts, photo-$z$ require $U$-band photometry to bracket the Balmer and 4000\AA\ breaks (e.g., \citealt{con95, saw97, Sorba2011}; see also Sec.~\ref{sec:photoz} of the  present paper).  Moreover, $U$-band is important for constraining star formation rates (SFRs) of galaxies at these redshifts, including through spectral energy distribution (SED) fitting \citep[e.g.,][]{saw98, sal09, sawicki2012sedfit}.  The photo-$z$ precision achievable with the CLAUDS+HSC-SSP data will allow rudimentary measures  of environment \citep[e.g.,][]{Malavasi2016, Moutard2018}, while the images are sufficient for basic morphological measurements [CLAUDS $U$-band data has median seeing of 0.92\arcsec (Sec.~\ref{sec:seeing}), and HSC $i$-band has 0.62\arcsec \citep{Aihara2019DR2}], and  identification of galaxy interactions using morphological features for which very deep images  are key \citep{Bridge2010}. 
  
The volumes probed by the CLAUDS+HSC-SSP data (e.g., $2\times10^7$ comoving Mpc$^3$ in the $z=0.6-0.8$ slice) approach those spanned by the SDSS main galaxy sample \citep{Strauss2002} and so contain  galaxy samples that are large enough to be split by environment, SFR, stellar mass, and morphology.  They are also large enough to overcome cosmic variance and will thus provide definitive measurements of the galaxy UV luminosity function (UVLF) and its evolution out to \zs3 via both photometric redshifts and, at \zs2--3, with BM/BX/$U$-drop Lyman Break Galaxy (LBG) techniques \citep[e.g., ][]{Steidel2004, saw06, hil09ii, saw12}. Using photometric redshifts or drop-out techniques, these data will also allow definitive measurements of galaxy clustering \citep[e.g.,][]{ade05, sav11} and other methods of linking galaxies and their dark matter halos, such as weak lensing and halo occupation distribution (HOD) modelling \citep[e.g.,][]{fou10, ouc04, cou12}. Using \zs2-3 BM/BX/LBG samples will permit magnification bias measurements of cluster masses at $z>1$, where traditional lensing is inadequate  \citep{bro95, hil09iii, Tudorica2017}. The $U$-band is also vital for providing the continuum measurement for line-emitters located in HSC's NB387 filter and creating large samples of $z=2.2$ Lyman-$\alpha$ emitters (LAEs). 
  
 The CLAUDS+HSC-SSP $z\sim2-3$ BM/BX/LBG and LAE samples will serve as low-redshift reference for higher-$z$ studies that use large areas and find rare, intrinsically luminous objects \citep[e.g.,][]{Konno2018, Matsuoka2018, Ono2018}.  The large areas covered by these data are also needed to find large samples of other intrinsically rare objects, including proto-clusters at \zs2--3 \citep[e.g.,][]{Toshikawa2016}, \zs3 Lyman Continuum Emitters \citep[e.g.,][]{iwata2009}, \zs3 quasars \citep[e.g., ][]{fon07}, Galactic white dwarfs and -- in combination with the NB387 filter -- several hundred  Lyman-$\alpha$ Blobs \citep[LABs; e.g., ][]{ste00, mat11}. Indeed, some such rare objects have already been found with these data, including Lyman continuum-emitting galaxies \citep{Bassett2019} and AGN (I.~Iwata et al., in preparation) at \zs3, and low-mass AGN at $z < 1$  (G.~Halevi et al., submitted; G.~Halevi et al., in preparation).

The present paper focuses on giving an overview of the CLAUDS project.  It contains a description of the CLAUDS fields and observations (Sec.~\ref{sec:fields-and-obs}), data reductions and an assessment of data quality (Sec.~\ref{sec:reductions-and-quality}), and several illustrative uses of the data (Sec.~\ref{sec:science-examples}): $U$-band galaxy number counts (Sec.~\ref{sec:numbercounts}), photometric redshifts (Sec.~\ref{sec:photoz}), and $U$-band drop-out selection (Sec.~\ref{sec:Udrops}). We use AB magnitudes throughout and assume the ($\Omega_M$, $\Omega_\Lambda$, $H_0$) = (0.3, 0.7, 70~km/s/Mpc) cosmology. We use \oldu\ and \newu\ to refer to the two specific MegaCam filters used in this project and \U\ to mean the general broadband wavelength region $\sim$3000--4000\AA\ or when we want to refer to the two filters, \oldu\ and/or \newu, without specifying either one in particular.   Table~\ref{tab:filters} gives some key details of the two $U$ filters (\newu\ and \oldu) used by CLAUDS, while their transmission curves are shown in Fig.~\ref{filters.fig}.

\section{CLAUDS fields and observations}\label{sec:fields-and-obs}

\subsection{Instrument and filters}

The instrument used for CLAUDS observations is MegaCam \citep{boul03}, a wide-field optical imager mounted on MegaPrime, the prime focus on the 3.6-metre CFHT.  While now surpassed in many respects by newer wide-field imagers such as HSC \citep{miyazaki2018} and DECam \citep{Flaugher2015}, MegaCam remains unrivalled in its sensitivity in the blue part of the spectrum. To take advantage of the excellent natural seeing at the CFHT site, MegaPrime includes a wide-field corrector that delivers uniform image quality over the entire field of view and an image stabilizing unit that effectively removes telescope wind shake.  Furthermore, just before the start of the CLAUDS observations, the CFHT enclosure was retrofitted with venting louvres that increase airflow through the dome \citep{Bauman2014} and are thought to improve seeing by $\sim$0.1\arcsec\ compared to previous CFHT MegaCam imaging programs such as the Canada-France Hawaii Telescope Legacy Survey \citep[CFHTLS; ][]{hud12} and the Next Generation Virgo Survey \citep[NGVS; ][]{fer12}.

MegaCam consists of 40 back-illuminated CCDs with good quantum efficiency in the near-UV, fabricated by e2v Technologies. Individual CCDs measure 2048$\times$4612 pixels; the pixels are 13.5 $\mu$m $\times$ 13.5 $\mu$m in size projecting to 0.187\arcsec $\times$ 0.187\arcsec on the sky. The 40 CCDs are arranged in a 4$\times$11 mosaic with the four corner positions empty.
The spacing between CCDs is approximately 13\arcsec, with larger gaps ($\sim$80\arcsec) between the uppermost and lowermost rows and the other CCDs. This configuration spans 1.21~$\deg \times$ 0.98~$\deg$, including the inter-chip gaps. A single exposure with all 40 CCDs produces a 378 megapixel image of 1.02 deg$^2$ (exposed sky area).  The CCDs take 40s to read out when --- as is standard --- two amplifiers are used per chip. The typical read noise is $\sim$5 e$^-$/pixel.  The cosmetic quality of the detectors is excellent and only $\sim$0.2\% of the pixels is unresponsive.

MegaCam is equipped with a filter jukebox that can accommodate up to eight filters.  However, the original MegaCam filters, $u^*g'r'i'z'$, used in previous programs such as CFHTLS, were only large enough to cover the central 9$\times$4 block of CCDs, thus illuminating 36 CCDs but leaving the outer four CCDs un-illuminated.  In late 2014 the filter set was upgraded with the purchase of new $ugriz$ filters.   These new filters have significantly better throughputs than the original filters.  The new filters are also physically larger, allowing all 40 CCDs of the array to be illuminated, instead of the 36 that were accessible by the old filters, thereby giving an $\sim$11\% improvement in collecting area.

\begin{figure}
\begin{center}
   \includegraphics[height=0.24\textheight]{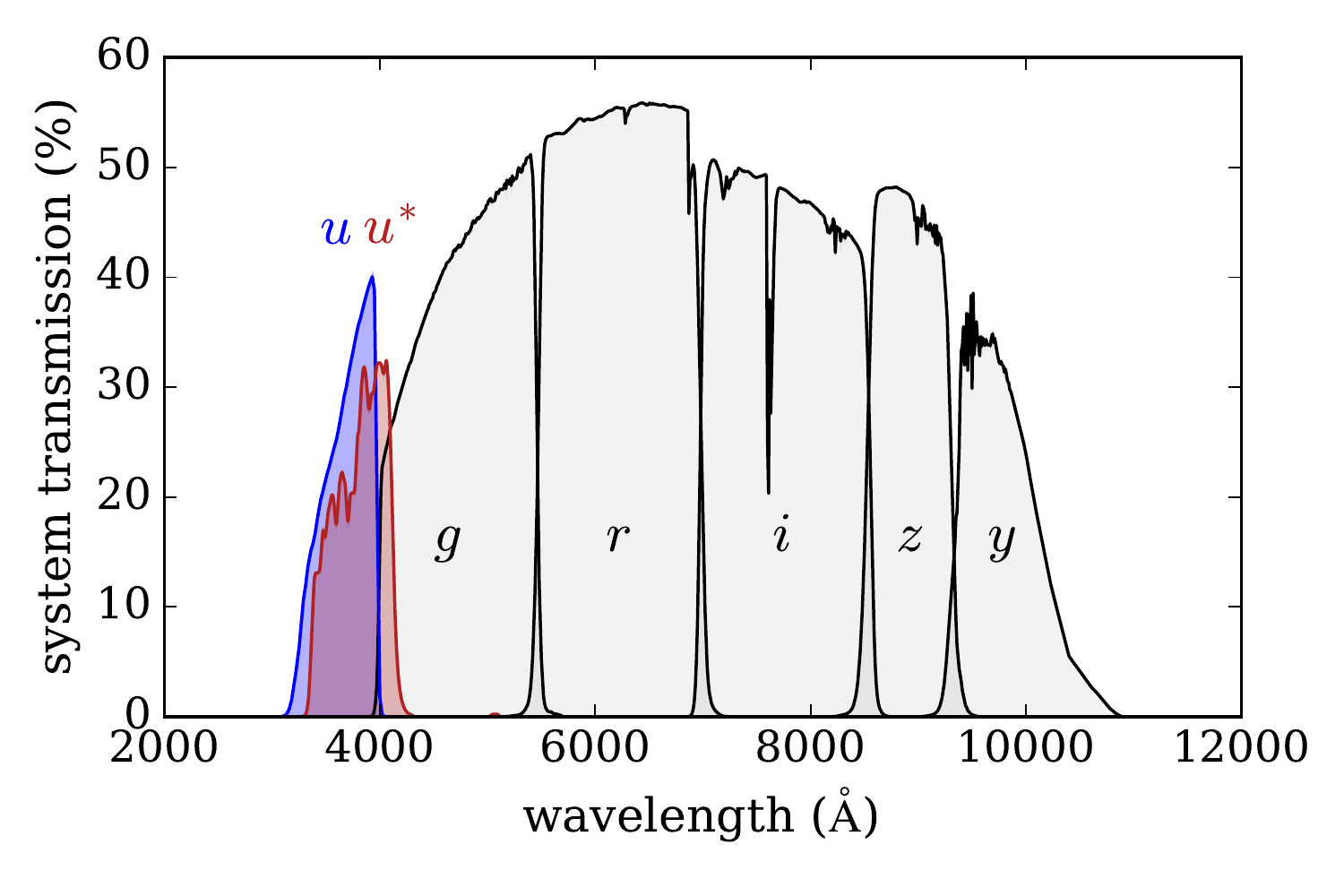}
\end{center}
 \caption[]{System transmission curves for the CLAUDS \newu\ and \oldu\ as well as the HSC $grizy$ bandpasses.  The filter transmission curves are taken from the CFHT website and also include the effects of telescope/instrument optics, CCD response, and 1.25 airmasses of typical Mauna Kea atmospheric extinction as described in \cite{Betoule2013}; the HSC filter transmission curves are from \cite{Kawanomoto2018} and also include telescope/instrument optics, CCD response, and 1.2 airmasses of extinction, all provided to us by the HSC-SSP collaboration. 
 Note that the new \newu\ filter (shown in blue) is significantly different from the old \oldu\ (red):  in addition to a higher throughput, its cut-on, central, and cut-off wavelengths are all bluer than those for the old \oldu. The new \newu\ filter also does not suffer from the old \oldu\ filter's red leak at $\sim$5000~\AA.}
\label{filters.fig}
\end{figure}

Of particular interest to our program are the $U$-band filters:  CLAUDS uses both the old \oldu\ filter and the new \newu\ filter (see Figure~\ref{filters.fig} for their filter transmission curves and Table~\ref{tab:filters} for some key details). CLAUDS consists of $U$-band data in four separate $\sim$5 deg$^2$ fields: E-COSMOS, ELAIS-N1, DEEP2-3, and XMM-LSS.  In two of these fields (ELAIS-N1 and DEEP2-3) CLAUDS data consist of \newu\ images only; one field (XMM-LSS) contains \oldu\ only; and one (E-COSMOS) has both \newu\ and \oldu.  The two filters, \oldu\ and \newu\ are sufficiently different that we treat them entirely separately, producing separate \newu\ and \oldu\ stacked images and object catalogues even in the $\sim$1~\sqdeg\ where we have overlap in \oldu\ and \newu\ imaging.

\begin{table}
\label{tab:filters}
\centering
\caption{
CLAUDS $U$ filters. Filter name is the name adopted by CLAUDS and follows the naming convention recommended by CFHT. Alternate name is the name used in some CFHT documentation; CLAUDS does not use these altrnate names, except for {\tt{uS}} which we use to mean \oldu\ in {\tt{ASCII}} file names and related data products.  CFHT ID is the number the filter is identified with in the CFHT filter database. Filter parameters are taken from the MegaCam filter database ({\tt {https://www.cfht.hawaii.edu/Instruments/Filters/megaprime.html}}) } 
\begin{tabular}{@{}lcccc@{}}
\hline
Filter  &  alternate  & CFHT ID & Central Wavelength & Bandwidth \\
name &  name  &  & (\AA) &  (\AA) \\
\hline
\newu\  & $u'$ & 9302 & 3538 & 868 \\
\oldu\ & $uS$, $U^\prime$ & 9301 & 3743 & 758 \\
\hline\
\end{tabular}
\label{fields.tab}
\end{table}

\begin{figure*}
\begin{center}
   \includegraphics[height=0.90\textheight]{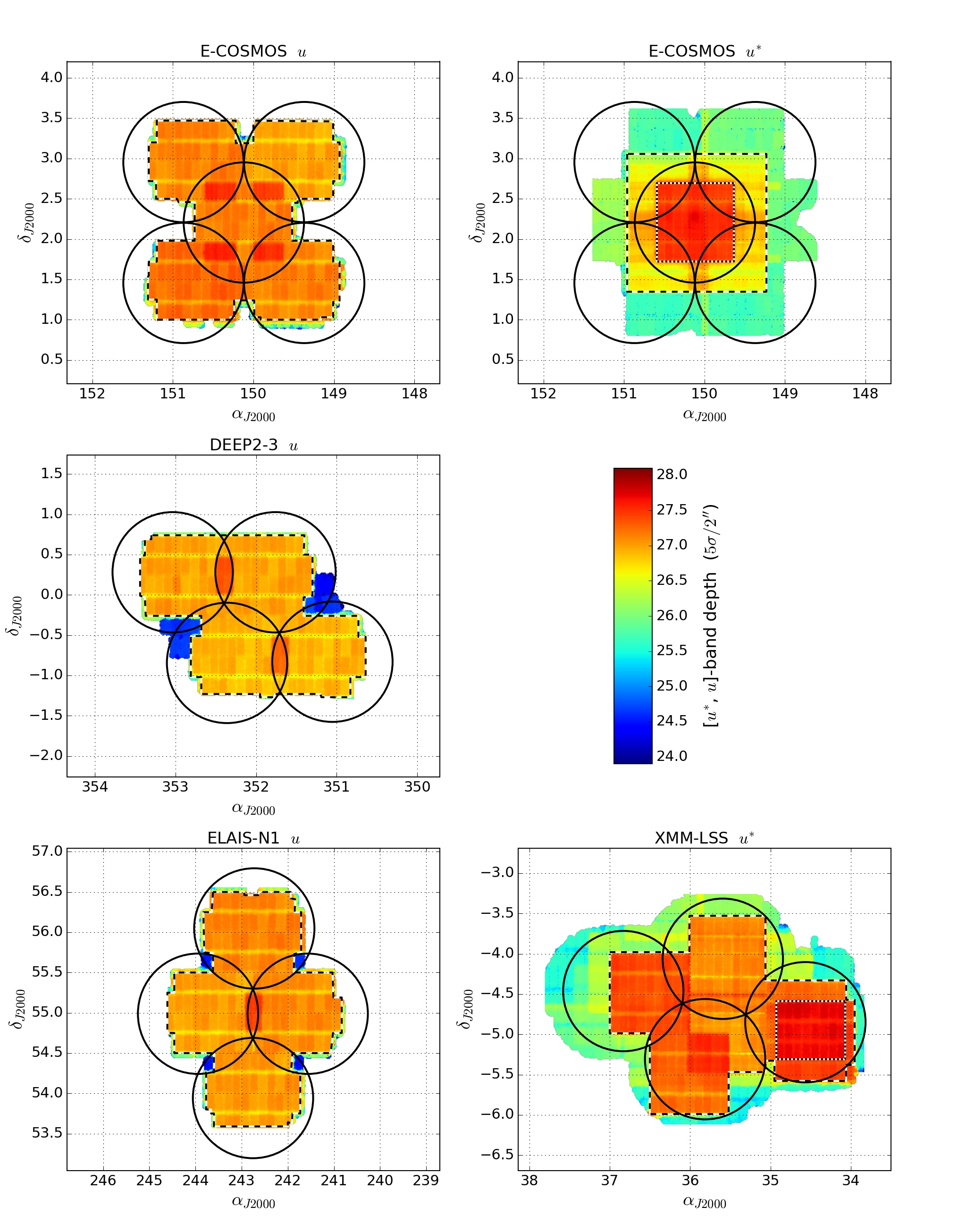}
\end{center}
 \caption[]{Depth and areal coverage of the CLAUDS data.  
The E-COSMOS field contains data in two $U$-band filters (\oldu\ and \newu) 
which are stacked separately; these are shown as two panels in this Figure. Black circles represent the nominal Subaru HSC pointings of the \hscs. Dashed lines mark the extent of the CLAUDS Deep data, and dotted lines that of the CLAUDS UltraDeep.  See \S~\ref{sec:depth} for details of the depth measurements.
}
\label{depthMaps.fig}
\end{figure*}

\subsection{The CLAUDS fields}

\begin{table*}
\centering
\caption{
CLAUDS fields and subfields.}
\begin{tabular}{@{}llcccl@{}}
\hline
Field \& subfield & other name &  RA centre (J2000) & Decl.\ centre (J2000) & filter & notes\\
\hline
XMM-LSS 0         &  SXDS &   02:18:15.60 &$-$04:51:00.0 & $u^*$ &   archival + MUSUBI data\\
XMM-LSS 1          &  &   02:22:11.00 &$-$05:00:00.0 & $u^*$ &  \\
XMM-LSS 2          &  &   02:22:11.00 &$-$04:03:00.0 & $u^*$ &  \\
XMM-LSS 3          &   &  02:25:59.00 &$-$04:29:40.0 & $u^*$ &  \\
XMM-LSS 4          &  &   02:24:03.00 &$-$04:29:40.0 & $u^*$ &  \\
XMM-LSS 5          & CFHTLS-D1 &   02:25:59.00 &$-$04:30:00.0 & $u^*$ &  archival data\\
E-COSMOS SW  &   &       09:58:03.00 &+01:27:21.0 & \newu &  \\
E-COSMOS NE   &   &      10:02:50.20 &+02:57:21.0 & \newu &  \\
E-COSMOS NW  &   &       09:58:03.00 &+02:57:21.0 & \newu &  \\
E-COSMOS SE   &   &      10:02:50.20 &+01:27:21.0 & \newu &  \\
E-COSMOS C     & CFHTLS-D2  / COSMOS &      10:00:28.60 &+02:12:21.0 & \newu & \\	
         \hspace{7mm} "     &      \hspace{14mm} "          &  "   &      "      & $u^*$ & archival + MUSUBI data  \\
ELAIS-N1 N         & &   16:11:00.00 &+56:00:00.0 & \newu &  \\
ELAIS-N1 S         &  &  16:11:00.00 &+54:00:00.0 & \newu &  \\
ELAIS-N1 W        &  &   16:07:35.00 &+55:00:00.0 & \newu &  \\
ELAIS-N1 E         &  &  16:14:25.00 &+55:00:00.0 & \newu & \\
DEEP2-3 SE        & &    23:28:47.00 &$-$00:45:55.0 & \newu & \\
DEEP2-3 NW       & &     23:27:37.00 &+00:13:50.0 & \newu & \\
DEEP2-3 SW       & &     23:24:53.00 &$-$00:45:55.0 & \newu & \\
DEEP2-3 NE        & &    23:31:31.00 &+00:13:50.0 & \newu & \\
\hline\
\end{tabular}
\label{fields.tab}
\end{table*}
CLAUDS consists of four fields  (the E-COSMOS, ELAIS-N1, DEEP2-3, and XMM-LSS fields), each covering $\sim$4--6~\sqdeg\  (Fig.~\ref{depthMaps.fig}) with several abutting or partially overlapping MegaCam subfields. The field locations (Table~\ref{fields.tab}) and layouts (Fig.~\ref{depthMaps.fig}) were chosen to maximally overlap with the $grizy$+NB imaging of the Deep Layer of the \hscs\ \citep{aih17overview} while respecting the on-the-sky footprints of both the HSC and MegaCam and, in some cases, the existence of archival data. In the ELAIS-N1 and DEEP2-3 fields all $U$-band data were taken by us with the new \newu\ filter. 

In the XMM-LSS field significant amounts of high-quality data taken with the old \oldu\ filter were available in the CFHT archive or --- while still proprietary at the time ---  were made available to us by the Mega-Cam Survey with $U$-band Imaging (MUSUBI; W.-H.\ Wang et al., in prep.).  To benefit from these data, we used the old \oldu\ filter to observe in this field.  In E-COSMOS, ELAIS-N1, and DEEP2-3 we used the \newu\ filter, taking advantage of its better throughput and area.  However, the central sqare degree of our E-COSMOS field, which corresponds to the intensively-studied COSMOS field \citep{scoville2007}, also has significant archival and MUSUBI \oldu-band data.  We reprocessed these \oldu\ data through our pipeline to produce a very deep \oldu\ stack.  Because we also observed this central area of E-COSMOS with the \newu\ filter, we have $\sim$1~\sqdeg\ of the sky with very deep data in both \newu\ and \oldu.  This overlap allows us to compare the \newu\ and \oldu\ data directly; it also increases the effective depth of $U$-band imaging available in this important and well-studied field.

\subsection{Observations}

CLAUDS is a program that combines time from three CFHT partners, namely Canada, France, and China.  Observing was done in queue mode over five semesters (2014B--2016B) with typical allocations of 80--86 hrs per semester with the exception of the final, fifth semester in which a 40 hr allocation was used to finish the remaining observations. In total, the CLAUDS project was allocated 376 hrs from the three Agencies; CFHT experience over many years shows that a MegaCam observing night results in 5.5 hours\footnote{http://cfht.hawaii.edu/en/science/LargePrograms/LP\_18B\_22A/ImplementationProcedures} of observing time on average (accounting for weather and telescope and instrument problems and averaging over the seasons and years), so these 376 allocated hours correspond to an allocation of $\sim$68 classical-mode observing nights. Moreover, because CLAUDS observing was usually allocated very high ranking in the observing queue, the data were taken in significantly better conditions (seeing, transparency) than would have happened in a  random set of 68 classical nights. 

Altogether, the dedicated CLAUDS observations resulted in useful images with a total open-shutter time of  $t_{exp}$= 280.60 hours (i.e., not counting overheads and weather or instrument problems); of this, 219.75 hours were taken with the new \newu\ filter, and 60.85 with the old \oldu. Additionally, we reprocessed and incorporated into our stacks 181.48 hours of archival \oldu\ data, including 51.62 hours of \oldu\ obtained in two ultra-deep $\sim$1-\sqdeg\ MegaCam pointings as part of the MUSUBI program.  In total, the full stacked images (dedicated CLAUDS plus archival data) represent 462.09 hours of open-shutter time, which is equivalent to $\sim$112 classical-mode nights (after accounting for overheads, telescope faults, and weather).

Dedicated CLAUDS observations were carried out in CFHT's queue mode during dark-time runs.  The bulk of the data was obtained using 600~s exposures taken in groups of either six or 11 dither positions (64 and 117 minute blocks of time, respectively, including overheads).   We used the pre-defined MegaCam Large Dithering Patterns LDP6 and LDP11, which cover a 30\arcsec$\times$180\arcsec ellipse\footnote{For dither positions and other details see \\ https://www.cfht.hawaii.edu/Instruments/Imaging/Megacam/specsinformation.html} and are designed to fill in the inter-chip gaps as well as to give good data for constructing sky flats and to ensure satisfactory masking of the few bad detector pixels. A very small number of shorter, 180~s exposures was also taken at positions that overlapped more than one of the main fields for the purpose of photometrically tying the fields together, if needed.  Where suitable, we also used archival MegaCam \oldu\ data -- including those from the MUSUBI program -- all retrieved from the CFHT archive at the Canadian Astronomy Data Centre\footnote{http://www.cadc-ccda.hia-iha.nrc-cnrc.gc.ca}  (CADC). Two subfields of the XMM-LSS field (see Table~\ref{fields.tab}) are based heavily on such archival data, while in the COSMOS field we used archival and MUSUBI data to construct the \oldu\ stacks. (The \newu\ data in E-COSMOS, and in the other fields, were all taken as new observations under the CLAUDS observing program.) The archival data were taken from programs executed between 2003 December 22 and 2016 March 31. These data were taken with a variety of observing strategies but with multi-point dither patterns and typical exposure times between 300--660 s (though a small number had shorter exposure times).

\section{Data reduction and quality}\label{sec:reductions-and-quality}

\subsection{Data reduction}

\subsubsection{Image processing, calibration, and matching with the \hscs\ data}
\label{sec:MegaPipe}
 
 As the first step in data reduction, the individual MegaCam images are pre-processed by the {\sc Elixir} software \citep{mag04} at CFHT. {\sc Elixir} applies detrending steps, namely overscan correction, bias subtraction, flat-fielding and masking, before the data are transferred to CADC for further processing.

Subsequent processing, namely astrometric/photometric calibration and image stacking, is done using the {\sc MegaPipe} data pipeline \citep{gwy08} at the CADC. The following is a summary of this {\sc MegaPipe} procedure, highlighting some of the modifications that were made to the original pipeline to accommodate the CLAUDS data.

Since the Gaia astrometric catalogue \citep{brown2016gaiaDR1} became available, {\sc MegaPipe} has used it as an astrometric reference frame.  However, early HSC astrometry, including that of Public Data Release 1 \citep[PDR1,][]{aih17DR1} and internal data relase S16A was tied to a pre-Gaia version of Pan-STARRS \citep{Magnier2016_PanSTARRS}, which in turn was tied to 2MASS \citep{Skrutskie2006_2MASS}.  In contrast, more recent versions of HSC-SSP data, including PDR2 \citep{Aihara2019DR2}, use Gaia astrometry.  Small but significant shifts, of the order of 0.1\arcsec, exist between these two reference frames and may be important for some science applications.  We therefore produce two versions of the CLAUDS image stacks, one matched to pre-Gaia Pan-STARRS astrometry, and one to Gaia astrometry. The MegaCam data used in Sec.~\ref{sec:science-examples} of the present paper use the pre-Gaia version of Pan-STARRS to match the reference frame of the HSC internal data release we used here.  The Gaia-calibrated CLAUDS stacks should, of course, be used with the more recent, Gaia-calibrated HSC images. The internal astrometric accuracy within the MegaCam data is better than 0.04\arcsec\ RMS, and the HSC and MegaCam images are aligned to each other with the same accuracy, provided that the pre-Gaia or post-Gaia stacks are used consistently. 

\begin{figure}
\begin{center}
\includegraphics[width=0.52\textwidth]{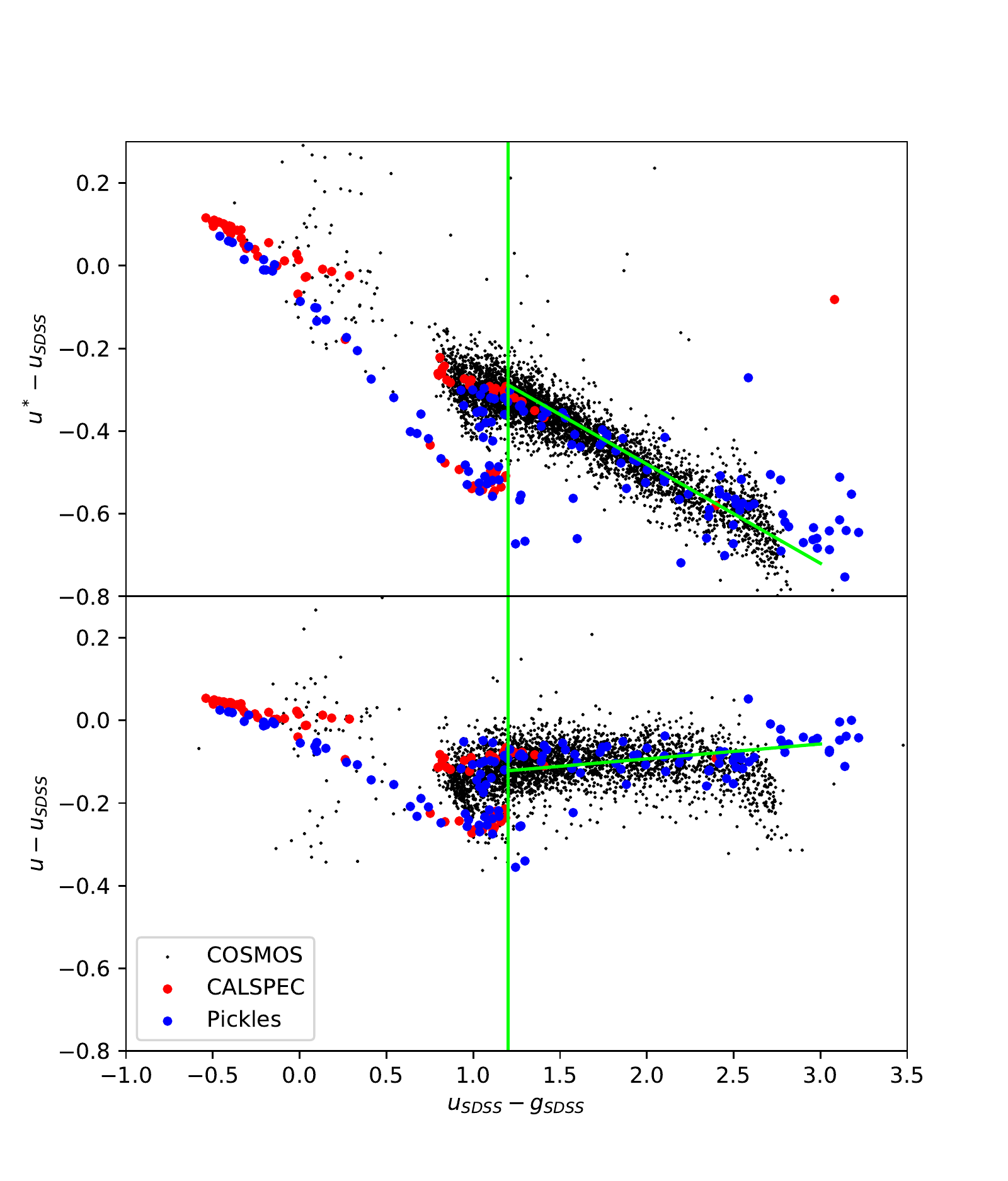}
\end{center}
\caption[]{ 
Transformations from the SDSS to the CFHT u-band filters.  Photometry from the COSMOS field is shown as black dots.  Synthetic photometry based on the CALSPEC \citep{Bohlin2014} and \citet{Pickles1998} spectra are shown in red and blue, respectively. The green line shows the adopted transformations between the SDSS and the CFHT $U$ filters, \oldu\ and \newu. The fits only take into account data redward of $u_{SDSS}-g_{SDSS}=1.2$, as indicated by the vertical line.  }
\label{ubandtrans.fig}
\end{figure}

The photometric calibration of CLAUDS is tied to the SDSS. The SDSS photometry is transformed into the the MegaCam system using the following transformations:
\begin{eqnarray}
\begin{aligned}
\label{eq:utrans}
u^*&= u_{SDSS}-0.241\ (u_{SDSS}-g_{SDSS})\\
u   &= u_{SDSS}+0.036\ (u_{SDSS}-g_{SDSS})-0.165, 
\end{aligned}
\end{eqnarray}
where \newu\ and \oldu\ are the CFHT filters while $u_{SDSS}$ and $g_{SDSS}$ refer to SDSS filters. These transformations are derived using synthetic photometry. To achieve this, the full response function of each filter is computed, including the transmission of the filter itself, the CCD quantum efficiency, the transmission of the MegaPrime optics, the reflectance of the CFHT primary mirror and 1.25 airmasses of atmospheric attenuation.  These response functions are multiplied by stellar spectra from \citet{Pickles1998} and \citet[the CALSPEC spectra]{Bohlin2014} -- the synthetic star colours thus produced are shown as blue and red points in Figure~\ref{ubandtrans.fig}. For comparison, colours of real stars in the COSMOS field are shown as black points. We ignore the effects of Galactic extinction since the differential extinction between the $U$ filters (\oldu/\newu\ and $u_{SDSS}$) is very small in this high-latitude fields. The fits given by Equations \ref{eq:utrans} are shown as a green lines. Note that the transformations are only valid for stars with  $u_{SDSS}-g_{SDSS}>1.2$.

When calibrating with the SDSS, we use only sources classified as stars by SDSS (using the SDSS parameter {\tt probPSF}).  We use the SDSS PSF magnitudes. The corresponding instrumental MegaCam magnitudes are measured through circular apertures whose size is set by the seeing of each image. Two apertures are used, with diameters 2 and 5.15 times the mean FWHM of stars in the image. The large aperture magnitude is equivalent to a Kron magnitude for point sources in MegaCam images. However, this aperture is quite large and consequently produces noisy measurements. The brighter stars in each image are used to determine an offset between the two sets of aperture magnitudes. The flux is then measured through the smaller aperture for all sources and corrected to the larger aperture. The zero-points computed in this way are useful for both stellar and galactic photometry.

For each MegaCam observing run, {\sc MegaPipe} builds a map of the differential zero-point offsets across the detector mosaic, using all the \newu\ (or \oldu) -band images which overlap the SDSS. This includes a large number of images that are not on CLAUDS fields, ensuring that the differential zero-point corrections are not affected by any local errors in the SDSS.  The differential corrections are computed on a CCD-by-CCD basis. This is adequate most of the time, with the caveat that for some runs there is some evidence that the four corner MegaCam chips have cross-chip gradients of about 1-2\%.  The differential corrections change somewhat between observing runs, with occasional larger changes due to changes in the {\sc Elixir} recipe.

Photometricity is determined using CFHT's {\sc SkyProbe} camera which monitors sky transparency once per minute by observing a set of standard stars located in the direction the telescope is pointing \citep{Cuillandre2002}. For each photometric night, {\sc MegaPipe} measures a nightly zero-point, again using all available images which overlap the SDSS. CLAUDS images taken on photometric nights are used to build a catalogue of photometric standards in the CLAUDS fields. This catalogue is then used to calibrate all the images, including those taken on non-photometric nights.

The photometric calibration of the individual images is cross-checked before stacking.  After the stacks are generated, their photometry is checked against that of the individual images. The calibration is found to be self-consistent to 0.005 magnitudes RMS. Both the input images and the stacks are compared back against the SDSS and we find that typical offsets are 0.015 magnitudes.  Note that although we use the SDSS as a calibration reference, we do not directly use the individual stars in the CLAUDS fields as standards, so this last test is effectively independent.

The astrometric/photometric calibrations computed above are stored in external header files. Once the input images are calibrated, they are stacked as follows, noting that \newu\ and \oldu\ data are stacked separately where they overlap (such as in the E-COSMOS field). To match the HSC pixel scale and image format, we produce stacks that cover exactly the 4200$\times$4100 pixel tiles corresponding to the "patches" in the \hscs\ data \citep{aih17DR1}; i.e., full mosaics covering the whole field are not constructed. The tiles measure 4200$\times$4100 0.168\arcsec $\times$0.168\arcsec\ pixels or about 0.2 degrees on the side.  To generate a tile, {\sc MegaPipe} determines which CCDs of which images overlap a particular patch. The CADC storage system allows individual FITS image extensions corresponding to the MegaCam CCDs to be extracted from the telescope archives with very little overhead. Therefore, only the CCDs relevant to a given patch are retrieved, greatly speeding the processing. These input images are resampled onto the HSC pixel grid using {\sc SWarp} \citep{ber02}
and a Lanczos-3 kernel.  Background subtraction is done with {\sc SWarp} using a 128$\times$128
pixel mesh. Finally, the images are combined by using a mean with 3$\sigma$ clipping.  Weight (inverse variance) maps for the images are also produced with the same format and pixel scale. The variance maps are produced by summing the variance maps of the individual input images, which in turn are derived from the sky in those images. 

The end product is a set of \newu\ and/or \oldu\ image tiles that have the same dimensions and pixel scale, and are astrometrically aligned with, \hscs\ $grizy$ image tiles. The images also have corresponding weight (inverse variance) maps.  As is the case for the HSC images, there is a small overlap between tiles which is useful in avoiding boundary effects in object detection and photometry.

\begin{figure*}
\begin{center}
   \includegraphics[height=0.49\textheight]{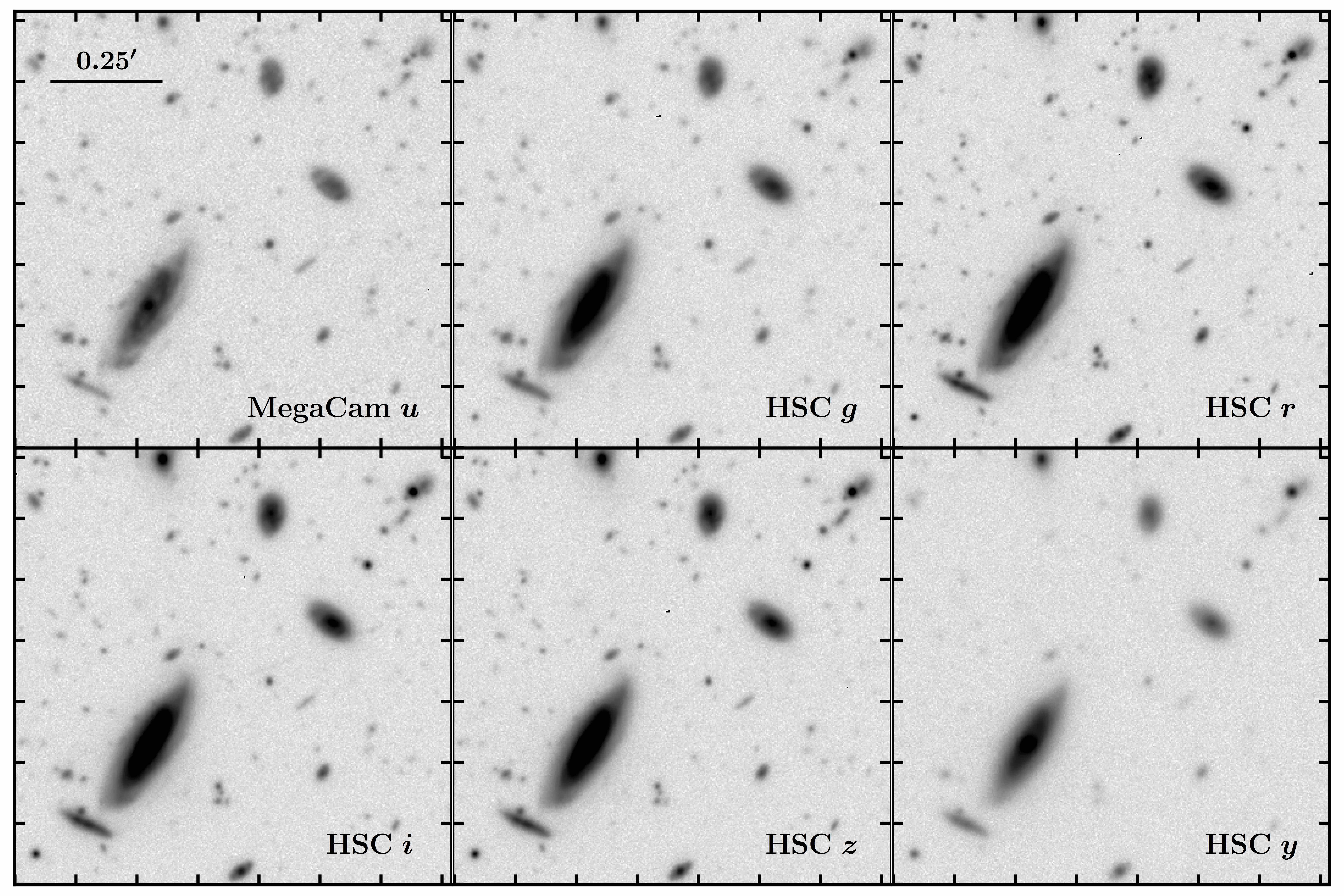}
\end{center}
 \caption[]{An example of the CLAUDS \newu\ and \hscs\ $grizy$ data at Deep depth.  Each panel shows the same representative 1'$\times$1' subfield within the central COSMOS field but imaged through a different filter. The \newu\ data in this field are of the depth that is typical of most of the CLAUDS survey ($u\approx 27$AB, 5$\sigma$ in 2\arcsec\ apertures), while the $grizy$ data shown here have the depth that will be typical of the \hscs\ Deep Layer once the \hscs\ is fully completed. When the HSC $grizy$ observations are completed, the combined surveys will contain 72,000 times the area shown in this 1'$\times$1' example, to comparable or deeper depths in $Ugrizy$. }
\label{dataExamples.fig}
\end{figure*}

To give a visual impression of the quality of both the CLAUDS and \hscs\ data, in Fig.~\ref{dataExamples.fig} we show a 1'$\times$1' subregion within the central COSMOS field.  Here, the \newu\ depth is typical of the entire CLAUDS dataset while the HSC $grizy$ images are at the depth that the whole \hscs\ Deep Layer will reach upon completion of Subaru observations, expected in 2021.

\subsubsection{Photometry and merged catalogues }\label{sec:photometry}

The steps described in \ref{sec:MegaPipe} give us \newu/\oldu\ images that are aligned with the \hscs\ $grizy$ images.  The next step is to produce combined multiband catalogues. For this, we use two completely independent, parallel procedures and produce two sets of independent catalogues. 

The first procedure applies the {\sc SExtractor} software \citep{bert96} to a merged $Ugrizy$ image constructed from combining the CFHT \newu/\oldu\ and Subaru $grizy$ images. The second procedure employs a modified version of the HSC data reduction pipeline \citep[{\sc hscPipe};][]{bos17}. Both procedures will be described in detail in forthcoming papers (A.\ Golob et al.\ in preparation; J. Coupon et al.\ in preparation), so here we only briefly describe their key elements. Having two independent photometric pipelines will let us in a future paper validate the relatively new {\sc hscPipe} applied to our $U$-band data against the simpler but well-established approach that {\sc SExtractor} represents. 

In both cases presented here we use our CLAUDS data as described in Sec.~\ref{sec:MegaPipe} and images from the S16A internal release of the \hscs\ data that are deeper than those from the first \hscs\ Public Data Release (PDR1, \citealt{aih17DR1}).  As the HSC data accumulate, future updates of our catalogues will incorporate newer HSC-SSP data releases. 

\begin{center}
{\emph {{\sc SExtractor}-based catalogues}}
\end{center}

We produce two catalogues using {\sc SExtractor}: (i) a single-band $U$ catalogue (keeping \newu\ and \oldu\ separate) and (ii) a multiband catalogue with object detection performed on a combined $Ugrizy$ image.

The $U$-band catalogue is created running {\sc SExtractor} on the \newu/\oldu\ images in the simple single-image mode with detection parameters tuned for this dataset. We record various measurements for each detected source, including position, fluxes (in Kron, isometric, and fixed-radius circular apertures), fiducial radius, ellipticity, position angle, and central surface brightness. 

For the multi-band catalogue, combined $Ugrizy$ images, $\Sigma \rm SNR$,  are created for each of the HSC patches by combining the observations in the $N$ bands observed as
\begin{equation}
\Sigma {\rm SNR}=\sum_{i=1}^{N}\left(\frac{f_{i}-\mu_{i}}{\sigma_{i}}\right),\label{eq:combUgrizy}
\end{equation}
where $f_i$ is the flux in each pixel, $\sigma_i$ is the RMS width of the background sky distribution, and $\mu_i$ is its mean. The index $i$ runs over MegaCam bands \newu\ or \oldu\ (or both, where available --- i.e.,  in the central area of E-COSMOS) as well as the HSC bands $grizy$.  
Like the $\chi ^2$ image \citep{Szalay1999} that's often used for object detection, the $\Sigma {\rm SNR}$ image combines information from all available bands. However, in contrast to $\chi^2$, $\Sigma {\rm SNR}$ places more weight on the shallower bands, which can improve the sensitivity to  objects with strong colors;  is less sensitive than $\chi^2$ to seeing variations between bands \citep[see][]{ber02}; and  avoids the possibility of spurious detections due to regions of correlated negative noise that can become positive when squared in the $\chi^2$ image.

The multiband catalogue is next created by running {\sc SExtractor} in dual image mode using the combined $Ugrizy$ $\Sigma \rm SNR$ images (Eq.~\ref{eq:combUgrizy}) for object detection.  For each detected source we record various measurements, including position, fluxes in all available bands (in Kron, isometric, and fixed-radius circular apertures), fiducial radii, ellipticities, position angles, and central surface brightnesses. We also record ``fluxes'' in the ``$\Sigma \rm SNR$'' band defined via Eq.~\ref{eq:combUgrizy}; these do not have a physical meaning, but are useful for understanding the detection properties of the catalogue. 

The \hscs\ consists of overlapping rectangular tracts which are divided into square patches. Adjacent HSC-SSP patches, as well as the CLAUDS data which are registered to them (\S~\ref{sec:MegaPipe}) overlap by 200 pixels so we discard all objects whose centres fall within 100 pixels of a patch edge, giving a catalogue of unique sources within a single tract. We merge the resulting tract catalogues by matching objects by position and, where duplicate objects are detected, keeping only the one with the highest S/N in the detection band (\oldu, \newu, or combined $Ugrizy$, depending on the catalogue in question).

\begin{center}
{\emph {{\sc hscPipe}-based catalogues}}
\end{center}

For the second procedure we use a modified version of the HSC pipeline \citep[{\sc hscPipe}; ][]{bos17} to combine the CLAUDS $U$-band data with the HSC $grizy$ data.  We modified {\sc hscPipe}, which is designed to work with HSC images, to allow it to handle our stacked MegaCam \newu/\oldu\ images.  All other features and functionality of {\sc hscPipe} remain unchanged. 

To adapt MegaCam data for ingestion into {\sc hscPipe}, we create a CLAUDS ``exposure'' object (in the form of a multi-extension FITS file) that contains the image, the variance map and the mask plane, all registered onto the same pixel grid and sky tiles (i.e., identical tracts and patches) as the HSC images. The variance map is simply the inverse weight map created by {\sc MegaPipe} (\S~\ref{sec:MegaPipe}) and the mask plane is identical to that of the $g$-band image, which contains the bright-star masks \citep{Coupon2018} and the geometry of the HSC-SSP survey.

To detect sources and accurately measure their photometry and morphology, the HSC pipeline accounts for point spread function (PSF) variations between the bands and across the image. The full software normally starts from single exposures, fits a spatially varying PSF model from high-SNR point sources and co-adds the PSF models in the same way as the exposures (for example, using a weighted-mean estimate). Here, for simplicity, we only measure the PSF on the co-added image. It is likely that the discontinuities of the PSF will not be properly reproduced in our best-fit PSF model, however, given the large number of co-added exposures in CLAUDS, we do not expect that this will have a significant impact on the measured photometry.

Next, we run the source detection on the \newu- and \oldu-band images only (this would be done during the co-addition process in the nominal version of the HSC pipeline), so that we can finally run the multi-band process on the full $Ugrizy$ dataset. This last step is done using the version of  {\sc hscPipe}  described by \citet{bos17}. In brief, {\sc hscPipe} first merges the footprints of the detections made on individual images, then picks the best-suited image to proceed with the de-blending of sources it considers to be blended  (the images are ordered per priority, the highest priority being the $i$-band image followed by other bands if the $i$-band SNR is too low), and finally performs a series of measurements (with uncertainty estimates) of the source photometry and morphology. Of particular note is that {\sc hscPipe}'s primary photometric output is CModel magnitudes for each band, including CLAUDS \newu\ and/or \oldu,  which are measured by finding the best PSF-convolved morphological model for each object (see \citealt{bos17} for details).

\subsection{Data quality}

\begin{figure}
\begin{center}
\includegraphics[width=0.52\textwidth]{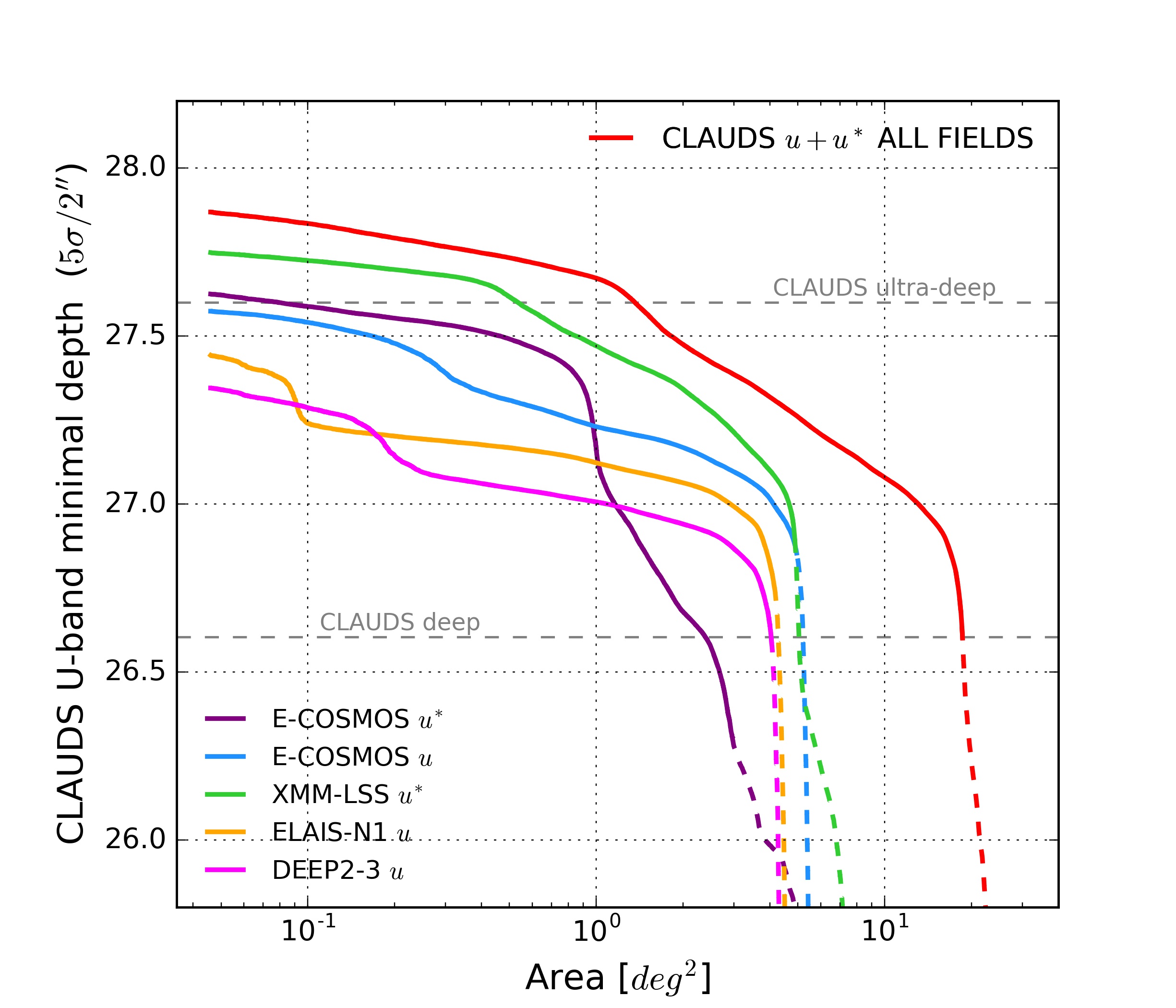}
\end{center}
\caption[]{ 
The limiting depth as a function of area for the total CLAUDS (red curve) and the individual constituent fields.  The \oldu\ and \newu\ are held separate for the E-COSMOS field.  The solid curves show the depth inside the footprints outlined in Fig.~\ref{depthMaps.fig}, while the dashed curves are outside those footprints. 
}
\label{CLAUDSlimitingDepths.fig}
\end{figure}

\begin{figure*}
\begin{center}
   \includegraphics[height=0.90\textheight]{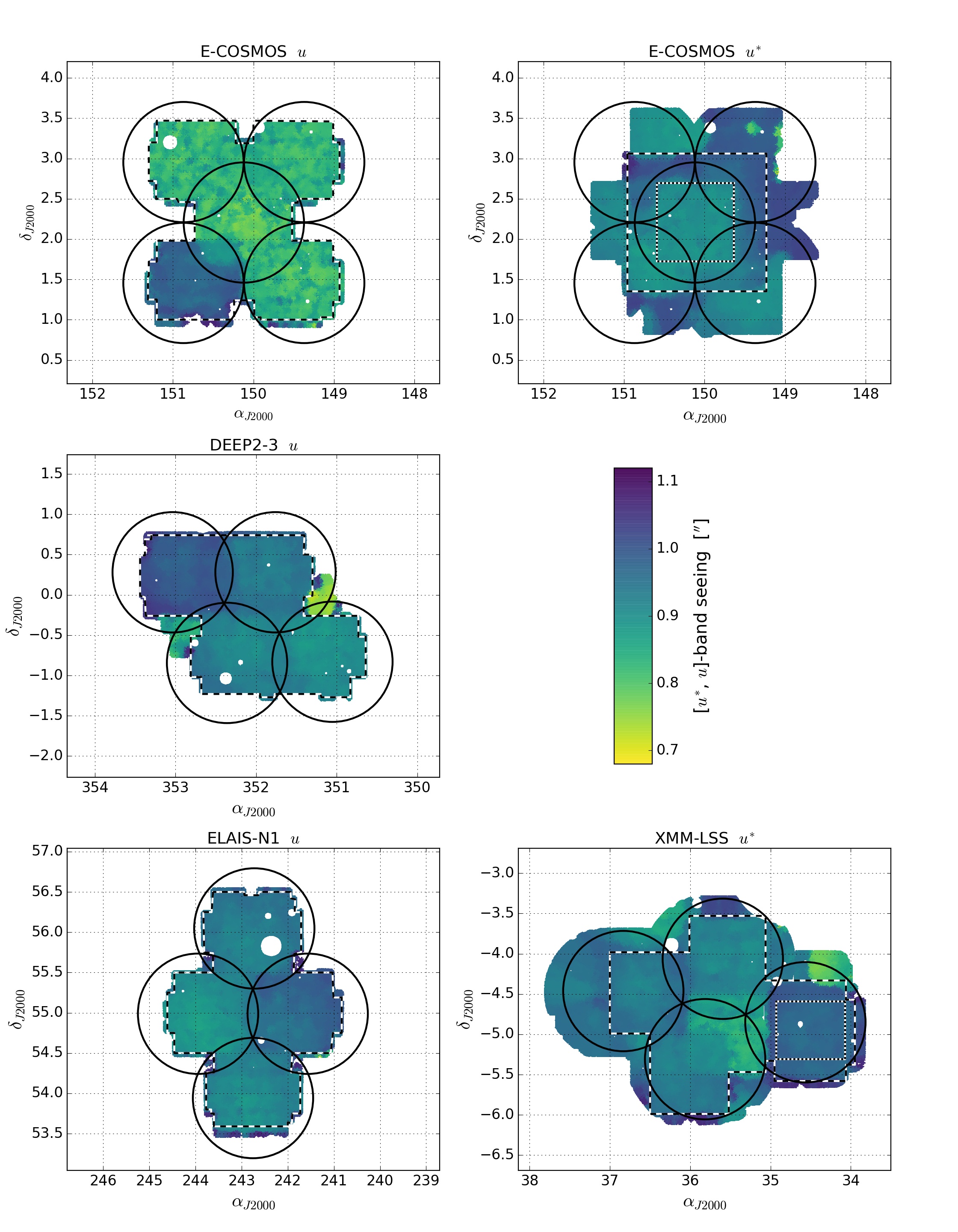}
\end{center}
 \caption[]{ Seeing in the stacked images measured from stellar half-light radii across the CLAUDS fields. Black circles represent the nominal Subaru HSC pointings of the \hscs.  The dashed lines mark the extent of the CLAUDS Deep data, and the dotted lines that of the CLAUDS UltraDeep. See \S~\ref{sec:seeing} for details of the seeing measurements. 
}
\label{seeingMaps.fig}
\end{figure*}

\subsubsection{Depth}\label{sec:depth}

Figure \ref{depthMaps.fig} shows the limiting magnitude (i.e., the depth) of our $u$- and $u^{*}$-band data across each CLAUDS field expressed as $5\sigma$ (i.e., SNR=5) in 2\arcsec-diameter apertures. This measurement was derived from the $U$-selected {\sc SExtractor}-based catalogue (Sec.~\ref{sec:photometry}) on a fine grid of positions by taking the median of the magnitudes of nearby sources that have SNR=5 in 2\arcsec-diameter apertures. This method has the advantage of providing empirical depth measurements with high angular resolution but requires reliable measurements of the photometric uncertainties.  The photometric uncertainties were calibrated empirically (see below) and this gave a multiplicative rescaling of the RMS by a factor of 1.5 to produce the final measurements shown in Fig,~\ref{depthMaps.fig}. 

\begin{figure*}
\begin{center}
   \includegraphics[height=0.45\textheight]{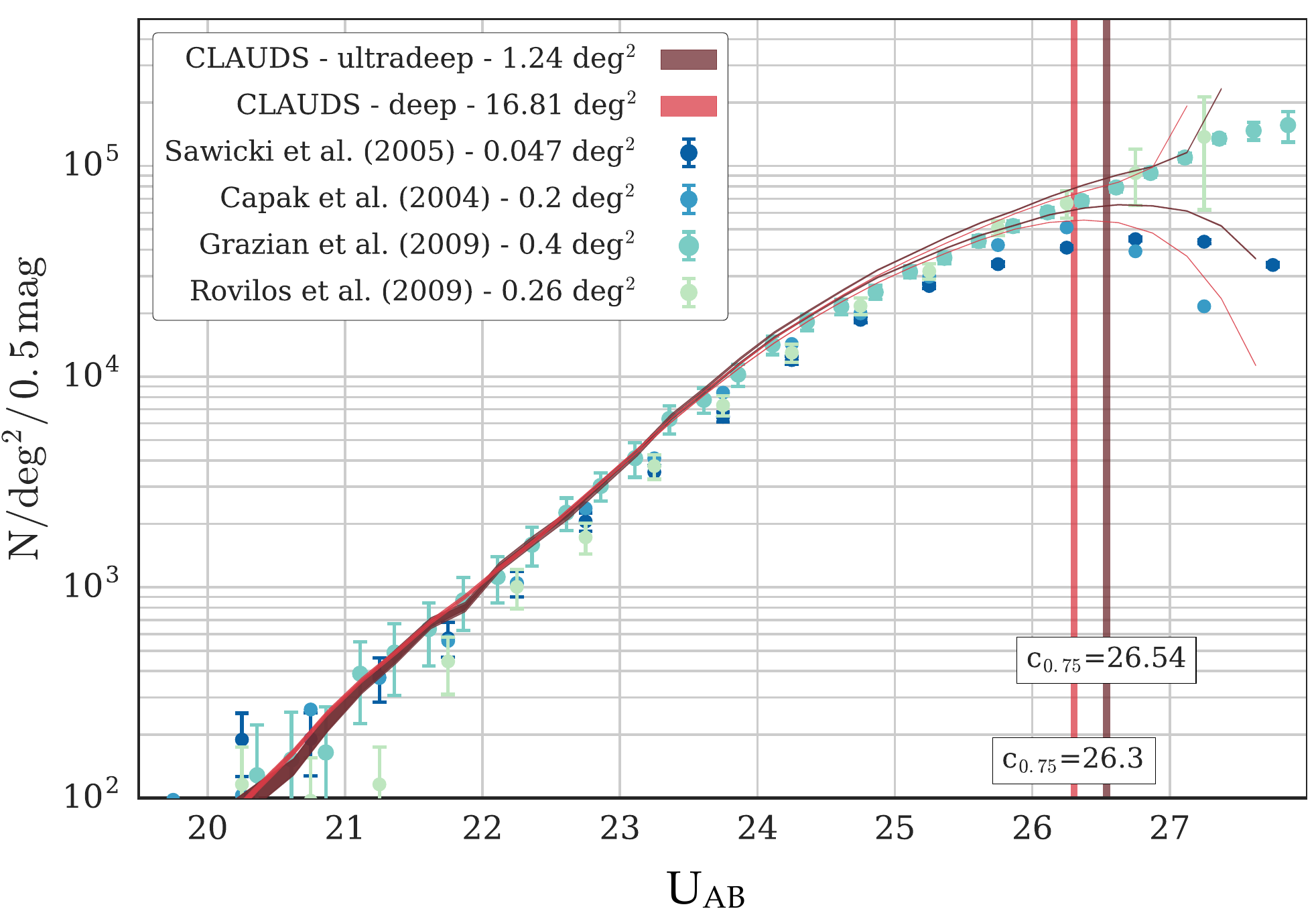}
\end{center}
 \caption[]{CLAUDS $U$-band galaxy number counts compared with other deep surveys. The CLAUDS counts are from {\sc SExtractor} object detection and photometry performed on $U$-band images only, as described in \S~\ref{sec:photometry}.  We found that there was no appreciably difference in the number counts in the \newu\ and \oldu\ bands and so we simply combine the data with an inverse-variance weighting. The CLAUDS counts are shown both in the Deep and UltraDeep, after excluding regions masked by bright stars or artifacts.  The Deep area listed does not include the UltraDeep area.   Both raw counts and incompleteness-corrected counts are shown: incompleteness-corrected counts are the higher two curves.  The width of the curves corresponds to 68~\% confidence regions and the vertical lines mark the 75\% completeness depths in the raw Deep and UltraDeep counts.  $U$-band counts from \cite{gra09} and \cite{rov09} are shown after correction for incompleteness by these authors, while the counts from \cite{cap04} and \cite{saw05} are raw, uncorrected counts. 
}
\label{numberCounts.fig}
\end{figure*}

The depth measurements shown in Fig.~\ref{depthMaps.fig}, were calibrated using two different estimates of the limiting magnitude, $m^{lim}$. In the first of these two approaches, we first created two images, each from an independent half of the available exposures (split by the MegaCam exposure number into  ``even" and ``odd" stacks with essential an equal number of individual exposures contributing to each stack).  The $m^{lim}$ was then derived from the difference of magnitude, $\Delta m$, measured for the same objects in the two half-exposure images.  Here, the depth of the half-depth stacks, $m^{lim}_{1/2}$, is obtained by comparing the RMS scatter in the magnitudes of objects detected at SNR=5. The full depth is then extrapolated through $m^{lim} = m^{lim}_{1/2} + 2.5 \log(\sqrt{2})$, where the $\sqrt{2}$ accounts for the difference in exposure time between the full and the half-exposure images. The second estimate of $m^{lim}$ was computed from the scatter of the background noise. To do so, the standard deviation of the background flux $\sigma_{bf}$ was measured in 2\arcsec\ apertures randomly placed across the image, while avoiding detected objects. The limiting magnitude is then given by $m^{lim} = -2.5 \log(\mathrm{SNR} \times \sigma_{bf}) + ZP $, where SNR = 5 and $ZP$ is the zero-point of the image. The two estimates of $m^{lim}$ agree with one another, after rescaling the raw $U$-band uncertainties by a factor of 1.5. Such uncertainty underestimates in the raw SNR measurements were expected, especially given the correlation between output pixels that is introduced by our stacking processes (see \S~\ref{sec:MegaPipe}). 

In Fig.~\ref{depthMaps.fig} the dashed lines outline an 18.60~\sqdeg\ CLAUDS Deep area outside of which the $U$-band depth falls off rapidly (see also~Fig.~\ref{CLAUDSlimitingDepths.fig}).  Within this Deep area, our data reach $U > 26.60$~AB  (5$\sigma$, 2\arcsec\ apertures) at every position, and the median depth is $U=27.09$~AB. The CLAUDS median depth is shown in Fig.~\ref{ubandSurveys.fig}, and the minimal depth as a function of area is shown in Fig.~\ref{CLAUDSlimitingDepths.fig}. Within the CLAUDS UltraDeep area, defined by the dotted lines in Fig.~\ref{depthMaps.fig}, our data reach $U > 27.60$ AB  (5$\sigma$, 2\arcsec\ apertures) at every position, and the median depth over this area is $U=27.71$~AB. 
	
To our knowledge, these are the deepest images ever taken in the $U$-band over such large areas.

\subsubsection{Seeing}\label{sec:seeing}

Fig.~\ref{seeingMaps.fig} shows $U$-band seeing maps measured from the stacked images.  To build these maps, we first extract from our $U$-selected {\sc SExtractor} catalogues a sample of high-probability stars with $U<24$ AB, S/N$>5$, and {\sc SExtractor} {\tt CLASS\_STAR}$>0.9$. At every position on the sky, we use the median of the half-light radii of all objects from this sample within a distance of 0.05 deg to characterize the local point spread function (PSF). Finally, we convert the half-light radii to the PSF FWHM using a conversion factor derived empirically from measurements of the MegaCam PSF \citep{gwy08}.  

The median seeing over the entire deep area of CLAUDS is 0.92\arcsec\, but varies across the survey, as can be seen in Fig.~\ref{seeingMaps.fig}. Nevertheless, it is typically $\la$1\arcsec\ over the survey and is substantially better in some subfields.

\section{Illustrative applications}
\label{sec:science-examples}

In this section we show three initial illustrative applications of our dataset: galaxy $U$-band number counts, $U$-band dropout number counts, and photometric redshifts.  All three results are preliminary and will be refined in separate papers (A.\ Golob et al., in prep; C.\ Liu et al., in prep).

\subsection{$U$-band galaxy number counts}\label{sec:numbercounts}

\begin{table*}
\centering
\caption{
CLAUDS raw $U$-band galaxy number counts for the Deep and UltraDeep regions of the survey.  Counts are raw counts (i.e., not corrected for incompleteness) and are given as  the logarithm of N deg$^{-1}$ 0.5 mag$^{-1}$. The 16--84 percentile uncertainty range is also given. The estimated completeness (columns 5 and 9) can be used to convert the raw counts to incompleteness-corrected counts.}
\begin{tabular}{@{}lcccccccc@{}}
\hline
U (AB) & counts & 16\% & 84\% & completeness & counts & 16\% & 84\% & completeness\\  
& log(N/deg$^2$/0.5 mag) & (Deep) & (Deep) & (Deep) & log(N/deg$^2$/0.5 mag) & (UD) & (UD) & (UD)\\
 & (Deep) & & & &(UD) &  & & \\
\hline
19.125 & 0.3059 & 0.1893 & 0.3976 & 1.0140 & --- & --- & --- & --- \\
19.375 & 0.8607 & 0.7997 & 0.9143 & 1.0118 & 0.8110 & 0.5100 & 0.9871 & 1.0144\\
19.625 & 1.2902 & 1.2544 & 1.3234 & 1.0093 & 1.1632 & 0.9871 & 1.3229 & 1.0144\\
19.875 & 1.5939 & 1.5696 & 1.6157 & 1.0065 & 1.6861 & 1.6041 & 1.7550 & 1.0141\\
20.125 & 1.8816 & 1.8629 & 1.8989 & 1.0019 & 1.8717 & 1.8000 & 1.9571 & 1.0116\\
20.375 & 2.0518 & 2.0382 & 2.0657 & 0.9994 & 2.0285 & 1.9724 & 2.0732 & 1.0070\\
20.625 & 2.2137 & 2.2021 & 2.2265 & 0.9991 & 2.1434 & 2.1066 & 2.1821 & 1.0042\\
20.875 & 2.4017 & 2.3928 & 2.4111 & 0.9988 & 2.3535 & 2.3156 & 2.3965 & 1.0014\\
21.125 & 2.5574 & 2.5498 & 2.5651 & 0.9965 & 2.5228 & 2.4964 & 2.5517 & 0.9968\\
21.375 & 2.6902 & 2.6834 & 2.6971 & 0.9940 & 2.6751 & 2.6400 & 2.7005 & 0.9943\\
21.625 & 2.8318 & 2.8259 & 2.8373 & 0.9914 & 2.8259 & 2.8033 & 2.8476 & 0.9937\\
21.875 & 2.9511 & 2.9459 & 2.9557 & 0.9892 & 2.9031 & 2.8838 & 2.9195 & 0.9912\\
22.125 & 3.0863 & 3.0822 & 3.0908 & 0.9886 & 3.0909 & 3.0739 & 3.1067 & 0.9863\\
22.375 & 3.2124 & 3.2089 & 3.2162 & 0.9863 & 3.2171 & 3.2005 & 3.2327 & 0.9815\\
22.625 & 3.3566 & 3.3536 & 3.3596 & 0.9838 & 3.3362 & 3.3236 & 3.3479 & 0.9790\\
22.875 & 3.4976 & 3.4951 & 3.5003 & 0.9813 & 3.4837 & 3.4733 & 3.4968 & 0.9787\\
23.125 & 3.6400 & 3.6378 & 3.6424 & 0.9787 & 3.6318 & 3.6226 & 3.6399 & 0.9784\\
23.375 & 3.7855 & 3.7835 & 3.7875 & 0.9762 & 3.8003 & 3.7949 & 3.8065 & 0.9759\\
23.625 & 3.9192 & 3.9176 & 3.9206 & 0.9736 & 3.9276 & 3.9223 & 3.9346 & 0.9711\\
23.875 & 4.0438 & 4.0423 & 4.0452 & 0.9708 & 4.0644 & 4.0588 & 4.0695 & 0.9660\\
24.125 & 4.1607 & 4.1594 & 4.1620 & 0.9657 & 4.1863 & 4.1820 & 4.1908 & 0.9609\\
24.375 & 4.2655 & 4.2645 & 4.2667 & 0.9580 & 4.2865 & 4.2827 & 4.2905 & 0.9555\\
24.625 & 4.3594 & 4.3585 & 4.3604 & 0.9481 & 4.3818 & 4.3786 & 4.3856 & 0.9476\\
24.875 & 4.4458 & 4.4449 & 4.4467 & 0.9376 & 4.4706 & 4.4683 & 4.4737 & 0.9351\\
25.125 & 4.5203 & 4.5194 & 4.5211 & 0.9249 & 4.5401 & 4.5380 & 4.5430 & 0.9201\\
25.375 & 4.5872 & 4.5865 & 4.5879 & 0.9093 & 4.6108 & 4.6084 & 4.6128 & 0.9045\\
25.625 & 4.6492 & 4.6485 & 4.6499 & 0.8881 & 4.6720 & 4.6694 & 4.6745 & 0.8864\\
25.875 & 4.7004 & 4.6998 & 4.7010 & 0.8544 & 4.7188 & 4.7168 & 4.7213 & 0.8632\\
26.125 & 4.7322 & 4.7316 & 4.7328 & 0.8040 & 4.7680 & 4.7661 & 4.7704 & 0.8318\\
26.375 & 4.7432 & 4.7426 & 4.7438 & 0.7389 & 4.8000 & 4.7979 & 4.8019 & 0.7865\\
26.625 & 4.7311 & 4.7305 & 4.7317 & 0.6451 & 4.8160 & 4.8141 & 4.8179 & 0.7298\\
26.875 & 4.6816 & 4.6809 & 4.6822 & 0.4795 & 4.8106 & 4.8085 & 4.8125 & 0.6543\\
27.125 &  ---   &  ---   &  ---   &  ---   & 4.7861 & 4.7842 & 4.7881 & 0.5156\\
\hline
\end{tabular}
\label{tab:Ucounts}
\end{table*}
 
Galaxy number counts are a standard way to characterize the observed galaxy population and also to test a new dataset and gauge its depth. The unprecedented combination of depth and area of CLAUDS allows a measurement of the $U$-band counts that has excellent statistics over a wide range in magnitude (see Fig,~\ref{numberCounts.fig}) and is largely insensitive to the effects of large scale structures (i.e., ``cosmic variance'').

Our CLAUDS number counts are derived using the single band (\newu- or \oldu-selected) {\sc SExtractor} catalogue. Effective areas are calculated by randomly sampling the image masks at one million points/deg$^2$. Stars were removed from the sample used to compute the number counts using {\sc SExtractor}'s {\tt CLASS\_STAR} parameter with a threshold of 0.85 and measured fluxes were corrected for Galactic extinction using the prescription of \cite{sch98}. $U$-band number counts and confidence intervals were computed for each area with 1000 bootstrap resamplings of the catalogue before a histogram was created. In each bootstrap iteration, the flux of each selected object was drawn from its Gaussian probability density function (PDF).  

The completeness of the catalogue was determined by adding synthetic objects to each of the detection images and measuring the recovery rate of objects as a function of flux. These artificial objects were generated using {\sc SExtractor}'s object-modelling functionality during the creation of the $U$-selected catalogues.  As such, each artificial object has a real counterpart in the $U$-band catalogues and has model properties (shape, size) that are representative of the real object. The additional crowding introduced by doubling the number of objects in the detection images could impact the completeness measurements; to minimize this bias, we record the distance of each real object in the original catalogue to its nearest neighbour and only use those {\sc SExtractor} model objects whose nearest neighbour in the combined (model+original) catalogue is no closer than the nearest neighbour of its original counterpart in the completeness calculations.

Table~\ref{tab:Ucounts} lists, for each magnitude bin, the median value of the 1000 bootstrap resamplings, along with the 16\% -- 85\% completeness intervals; the values are shown graphically in Fig. \ref{numberCounts.fig} which shows the number counts calculated over the CLAUDS Deep and UltraDeep areas (red and brown shaded regions respectively) and their incompleteness-corrected counterparts.  Stellar objects have been removed and magnitudes for the CLAUDS number counts are {\sc SExtractor}'s {\tt MAG\_AUTO} measurements.  Note that the CLAUDS Deep and UltraDeep areas given in Fig.~\ref{numberCounts.fig} are after correcting for regions masked due to bright stars and artifacts.  

We compare our $U$-band counts to previous deep $U$-band surveys (all taken with 8-metre-class telescopes: \cite{cap04}, \cite{saw05}, \cite{gra09}, \cite{rov09}) and find our result to be in good agreement with these earlier measurements. CLAUDS 75\% $U$-band detection completeness, determined from our simulations, is shown with vertical lines in the Figure and is $c_{0.75}=26.30$~AB in the CLAUDS Deep area and $c_{0.75}=26.56$~AB in the UltraDeep.

\subsection{Photometric redshifts}\label{sec:photoz}

\begin{figure}
\begin{center}
   \includegraphics[height=0.55\textheight]{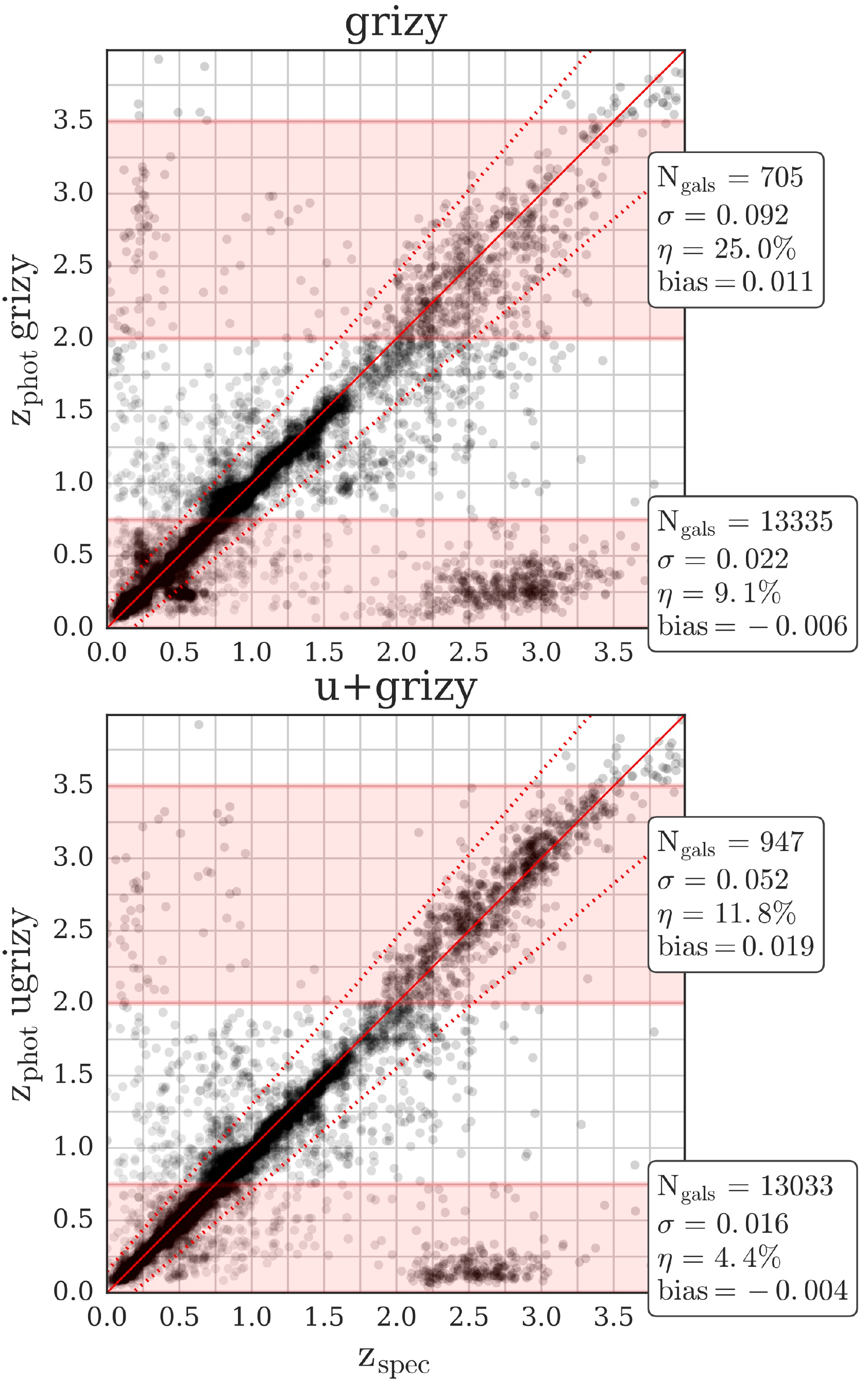}
\end{center}
 \caption[]{Photometric redshifts without {\it(top)} and with {\it (bottom)} $U$-band.   The addition of $U$-band (in this case \oldu) to the $grizy$ dataset significantly improves photometric redshift performance.  Scatter ($\sigma$, defined as the normalized median absolute deviation, NMAD), outlier fraction ($\eta$) and bias are all strongly reduced, particularly at $z < 0.75$ and $2<z < 3.5$.  The statistics presented in the figure are for the two highlighted bands in $z_{phot}$. The bias at $z\sim 2.5$ in the bottom panel is due to a similar bias that our empirical method inherited from the \cite{lai16} photometric redshifts that we use as our training set; additionally, many of the catastrophic outliers at $z_{spec}\sim2.5$ in the bottom panel are due to insufficient training and with further refinements of the method (in prep.) we can eliminate $\sim$75\% of them.
}
\label{photo-z.fig}
\end{figure}

\begin{figure*}
\begin{center}
   \includegraphics[width=1.0\textwidth]
{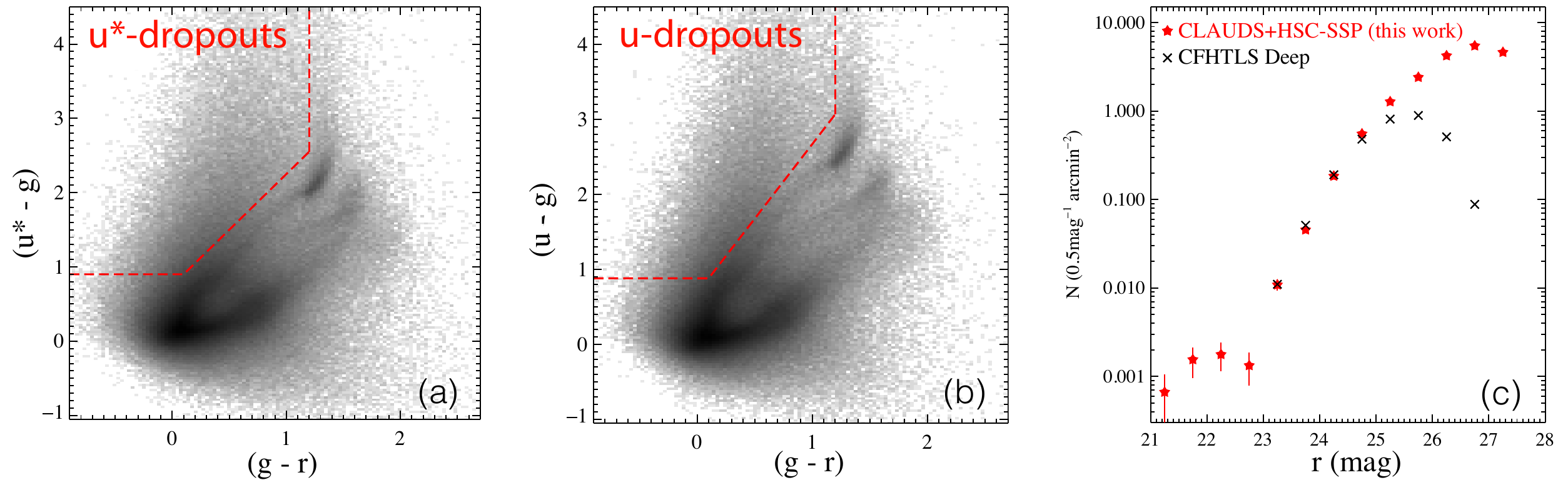}
\end{center}
 \caption[]{Colour-colour selection of \oldu-band dropouts (left panel) and \newu-band dropouts (middle panel) overlaid on grayscale density plots of objects in the CLAUDS+HSC-SSP. Right panel: number counts of CLAUDS $U$-band dropouts (combined \oldu- and \newu-dropouts). Our data already allow us to go considerably deeper in $U$-band dropout studies and over a considerably larger area than was possible in the CFHTLS Deep \citep[black crosses: ][]{hil09ii}. }
\label{Udrops.fig}
\end{figure*}

Given the depth and large area of CLAUDS and HSC-SSP imaging, spectroscopy will not be possible for the vast majority of objects even  in the eras of the Prime Focus Spectrograph \citep[PFS;][]{tak14} now under construction for Subaru and  the planned Maunakea Spectroscopic Explorer \citep[MSE;][]{McConnachie2016}. Quality photometric redshifts are therefore essential for many science applications. Photometric redshifts rely on the detection of prominent spectral breaks in multi-band photometric measurements, and here $U$-band fluxes are essential for localizing the Balmer and 4000\AA\ breaks at lower redshifts, and the Lyman break at higher redshifts \citep[e.g.][]{con95, Gwyn1996, saw97}.  In this section we illustrate how the combination of CLAUDS $U$  and HSC-SSP $grizy$ observations can give excellent photometric redshifts. 

Our team is using several photometric redshift techniques and these will be described in future papers. For the illustration in the present paper, 
 our photometric redshifts are computed using a colour-space nearest-neighbour machine learning technique that will be described in a forthcoming paper (A.\ Golob et al., in preparation).  Briefly, we use the 30-band COSMOS photometric redshifts from \cite{lai16} as a training set.  We match our {\sc SExtractor}-based $uu^*grizy$ catalogue (\S~\ref{sec:photometry}) to that of \cite{lai16} by object positions with a tolerance of 1\arcsec.   For each object in our full catalogue, we calculate colours from {\tt {MAG\_ISO}} photometry and identify the 50 nearest neighbours in colour space with matched COSMOS photo-z's.  We use these 50 nearest neighbours to fit a weighted Gaussian kernel density estimator (KDE), with each neighbour's redshift weighted by $(d_\mathrm{NN}\times \Delta z)^{-1}$, where $d_\mathrm{NN}$ is the Euclidean distance in feature (i.e., colour) space to the object under consideration, and $\Delta z$ is the width of the 68\% confidence interval of the neighbour's redshift in the \cite{lai16} catalogue. Where $d_\mathrm{NN}=0$, we set weights to 0 to prevent objects in the COSMOS field from using {\it{themselves}} in their KDEs.

Adding the $i$-band magnitude to the feature space used to select each object's nearest neighbours effectively acts as a prior probability on the object's redshift: when it is included, objects with brightnesses that differ significantly from the object under consideration are moved to larger distance and penalized in the KDE if included. This is helpful in minimizing outliers at $z_{spec} \sim 3$, but worsens the performance of the estimator at low redshift.

\cite{Tanaka2018photoz} presented photometric redshifts computed from \hscs\ $grizy$ (but no $U$) data using several techniques.  However, we do not compare our photometric redshifts with their results as our goal here is to test how the addition of $U$-band improves photometric redshifts.  To do so, we should use the same photometric redshifts code on catalogues that do or do not include the $U$ band.  We thus compare photometric redshift results produced by our machine learning procedure applied to $grizy$ data alone and to $grizy$ combined with $U$-band.  We do the comparison in the E-COSMOS C field where we have both \oldu\ and \newu\ data (which we treat as separate measurements), although the two filters are sufficiently similar to each that we are justified in treating this test as one that tests photo-$z$ improvement brought about by the addition of any deep $U$-band to the \hscs\ broadband data. 

As can be seen in Fig.~\ref{photo-z.fig}, the inclusion of the $U$-band data dramatically improves the quality of photometric redshifts. In this Figure we compare our photometric redshifts calculated without $U$-band (top panel) and with (bottom panel) against 22,005 high-quality spectroscopic redshifts we extracted from a compilation of surveys in our fields \citep[][and D. Masters, private communication]{bra13, com15, lef13, kri15, lil07, mas17, mcl13, sco18, sil15, tas17}. Of these redshifts, 13,779 have $z_{spec}<0.75$ (604 of these are stars and were excluded from the sample) and 1,213 are at $2<z_{spec}<3.5$.  It is clear that the addition of $U$-band improves photometric redshift performance, particularly at $z_{phot}<0.75$ and $2<z_{phot}<3.5$ (highlighted bands in the Figure).

To quantify the improvement due to the addition of $U$-band data, we calculate the scatter, $\sigma$, the catastrophic outlier fraction, $\eta$, and the bias in our photometric redshifts. We define $\sigma$ as the normalized median absolute deviation (NMAD)$: \sigma=1.48\times \mathrm{median}(\Delta z/(1+z_{\mathrm{spec}}))$, where $\Delta z$ is the absolute difference between  $z_\mathrm{spec}$ and $z_\mathrm{phot}$. The outlier fraction $\eta$ is defined as the fraction of galaxies with $\Delta z>0.15 \times (1+z_\mathrm{spec})$. Finally, the bias is defined as the median value of $(z_\mathrm{phot}-z_\mathrm{spec})/(1+z_\mathrm{spec})$. 

The addition of $U$ to the $grizy$ gives improvements in photo-$z$ performance over the full redshift range.  Over the full redshift range, the addition of $U$-band reduces scatter from $\sigma_{grizy}=0.023$ to $\sigma_{Ugrizy}=0.020$, the bias from $-0.006\times (1+z)$ to $-0.004\times (1+z)$, and the outlier fraction from $\eta_{grizy}=5.3\%$ to $\eta_{Ugrizy}=5.2\%$.  The improvements are particularly impressive at $z_{phot}<0.75$ and $2<z_{phot}<3.5$. At $z_{phot}<0.75$, adding the $U$ band reduces the measured scatter from $\sigma_{grizy}=0.022$ to $\sigma_{Ugrizy}=0.016$ and the outlier fraction from $\eta_{grizy}=9.1\%$ to $\eta_{Ugrizy}=4.4\%$ with negligible change to the bias. At $2<z_{phot}<3.5$ the scatter drops from $\sigma_{grizy}=0.092$ to $\sigma_{Ugrizy}=0.052$ and the outlier fraction drops from $\eta_{grizy}=25\%$ to $\eta_{Ugrizy}=11.8\%$ while the bias increases from 0.011 to 0.019, reflecting systematic errors inherited from the \cite{lai16} photometric redshifts that we used as our training set.   These improvements are fully consistent with the fact that photometric redshifts rely on the straddling of spectral breaks with photometric bands:  as is well known, $U$ fluxes are needed to straddle the Balmer and 4000\AA\ breaks at lower redshifts, and the Lyman break at higher redshifts \citep[e.g.][]{con95, Gwyn1996, saw97}.

We note that our empirical photo-$z$ technique is trained using the \cite{lai16} photometric redshifts which in turn use the {\sc LePhare} photo-$z$ code \citep{Arnouts2002} calibrated using spectroscopic redshifts.  Many of the same spectroscopic redshifts also feature in the comparisons shown in Fig.~\ref{photo-z.fig} and this might lead one to suspect there is a circularity involved in our photo-$z$ performance statistical measures.  However, the LePhare photo-$z$ calibration is very simple: it is only used to adjust their photometry zeropoints globally.  Therefore, this means that there are no biases  in our photo-$z$ statistics introduced by this approach; moreover, the improvement in $Ugrizy$ over $grizy$ photometric redshift sets is real as both sets are calculated with the same procedure and spectroscopic samples.

\subsection{$U$-band Drop-outs at \zs3}\label{sec:Udrops}

The Lyman break technique \citep{Guhathakurta1990, Steidel1996} gives a well-established way to identify and study large samples of star-forming Lyman break galaxies (LBGs) at high redshift \citep[e.g.][and many others]{Steidel1998, Steidel1999, Sawicki2006b, hil09ii, Ono2018}.  Selection of \zs3 LBGs, or $U$-band drop-outs, requires  very deep $U$-band photometry as well as deep data at two longer wavelengths \citep{Steidel1996}, and the combination of CLAUDS and HSC-SSP are excellent for this purpose. Here, the large area of CLAUDS+HSC-SSP will yield very large samples of \zs3 LBGs that will allow not just excellent statistics and insensitivity to cosmic variance, but will also probe a rich variety of environments \citep[e.g.][]{Toshikawa2016}. In this section we illustrate the ability of CLAUDS $U$ data, in conjunction with $gr$ photometry from HSC-SSP, to select \zs3 $U$-band dropout LBGs. 

We select Lyman break galaxies at $z\sim3$ using our {\sc hscPipe}-produced catalogues (see \S~\ref{sec:photometry}).  We use {\sc hscPipe} catalogues here for commonality with \hscs\ studies of LBGs at higher redshifts \citep[e.g.,][]{Ono2018}. Because CLAUDS consists of \oldu\ and \newu\ observations in different fields, we used both these filters for our dropout selection by employing two similar but different selection criteria for \oldu-droputs and \newu-dropouts. 

For the $u^*$-dropout selection, we use selection criteria similar to those in \cite{hil09ii} who used the very same MegaCam \oldu filter, and MegaCam $g'$ and $r'$ filters that are similar to the HSC $g$ and $r$.  The colour selection in the $u^*gr$ diagram is described as 
\begin{equation} \label{eq:uSgr}
\begin{split}
u^*-g & > 0.9,\\
g-r   & < 1.2,\\
u^*-g & > 1.5(g-r)+0.75, 
\end{split}
\end{equation}
and is shown with the red lines in the left panel of Fig.~\ref{Udrops.fig}.

As can be seen in Figure~\ref{filters.fig}, the central wavelength of the new $u$-band filter is bluer than that of the old $u^*$ filter.  To account for this difference, we modified the \oldu-dropount selection criteria of Eq.~\ref{eq:uSgr} and
define the selection window in the $ugr$ diagram as
\begin{equation} \label{eq:ugr}
\begin{split}
u-g & > 0.88,\\
g-r   & < 1.2,\\
u-g & > 1.88(g-r)+0.68.
\end{split}
\end{equation}
This selection window is shown with the red lines in the middle panel of Fig.~\ref{Udrops.fig}. 

We combine our \oldu-dropout and \newu-dropout number counts and plot them in the right panel of Fig.~\ref{Udrops.fig}. Our results (red symbols with error bars) are comparable to previous results in the CFHTLS-Deep \citep[black crosses,][]{hil09ii} at intermediate magnitudes ($r\sim23-25$). However, at fainter magnitudes it is clear that our sample goes deeper than this previous work; it will go deeper still once the HSC-SSP imaging reaches its full depth at the conclusion of the survey. It is also worth emphasizing that our sample covers significantly more area than the $\sim$4~deg$^2$ CFHTLS-Deep work of \citet{hil09ii}. We will present $z\sim3$ LBG luminosity and correlation functions in forthcoming papers (C.~Liu et al., in prep.\ and Y.~Harikane et al., in prep).

\section{Outlook}

The CLAUDS $U$-band data are all in hand and combined with the existing HSC-SSP \grizy\ dataset.  We are now  using these data to tackle a number of scientific studies, and several papers are in preparation. 

While the CLAUDS $U$-band data acquisition is complete, the HSC-SSP project is ongoing, with more exposures being acquired in the fields covered by CLAUDS.  We plan to regularly update our catalogues to reflect this growing $grizy$ depth and expect to publicly release out dataset -- both catalogs and images -- in 2020.

\section*{Acknowledgments}

We thank the CFHT observatory staff for their hard work in obtaining these data. The observations presented here were performed with care and respect from the summit of Maunakea which is a significant cultural and historic site.  We also thank the expert anonymous referee who suggested a number of important details to be included in the paper. 

This work is based on observations obtained with MegaPrime/MegaCam, a joint project of CFHT and CEA/DAPNIA, at the Canada-France-Hawaii Telescope (CFHT) which is operated by the National Research Council (NRC) of Canada, the Institut National des Science de l'Univers of the Centre National de la Recherche Scientifique (CNRS) of France, and the University of Hawaii. This research uses data obtained through the Telescope Access Program (TAP), which has been funded by the National Astronomical Observatories, Chinese Academy of Sciences, and the Special Fund for Astronomy from the Ministry of Finance. This work uses data products from TERAPIX and the Canadian Astronomy Data Centre. It was carried out using resources from Compute Canada and Canadian Advanced Network For Astrophysical Research (CANFAR) infrastructure.  We are grateful for research support funding from the Natural Sciences and Engineering Research Council (NSERC) of Canada.

This work is also based in part on data collected at the Subaru Telescope and retrieved from the HSC data archive system, which is operated by the Subaru Telescope and Astronomy Data Center at National Astronomical Observatory of Japan. The Hyper Suprime-Cam (HSC) collaboration includes the astronomical communities of Japan and Taiwan, and Princeton University.  The HSC instrumentation and software were developed by the National Astronomical Observatory of Japan (NAOJ), the Kavli Institute for the Physics and Mathematics of the Universe (Kavli IPMU), the University of Tokyo, the High Energy Accelerator Research Organization (KEK), the Academia Sinica Institute for Astronomy and Astrophysics in Taiwan (ASIAA), and Princeton University.  Funding was contributed by the FIRST program from Japanese Cabinet Office, the Ministry of Education, Culture, Sports, Science and Technology (MEXT), the Japan Society for the Promotion of Science (JSPS),  Japan Science and Technology Agency  (JST),  the Toray Science  Foundation, NAOJ, Kavli IPMU, KEK, ASIAA,  and Princeton University.This paper makes use of software developed for the Large Synoptic Survey Telescope. We thank the LSST Project for making their code available as free software at http://dm.lsst.org. 

During the course of this work, we have made use of the cosmological calculator of \citet{Wright2006}.






\bibliographystyle{mnras}
\bibliography{CLAUDS-1.bib}



%
%

\appendix
\section{Affiliations}
\label{sec:affiliations}

$^{1}$Institute for Computational Astrophysics \& Department of Astronomy and Physics, Saint Mary's University, Halifax, Canada\\
$^{2}$Aix Marseille Universit\'e, CNRS, Laboratoire d'Astro\-phy\-sique de Marseille, UMR 7326, F-13388,  Marseille, France\\
$^{3}$CASSACA, National Astronomical Observatories of China\\
$^{4}$Harvard-Smithsonian Centre for Astrophysics, USA\\
$^{5}$Astronomy Department, University of Geneva, Chemin d'Ecogia 16, CH-1290 Versoix, Switzerland\\
$^{6}$NRC-Herzberg, 5071 West Saanich Road, Victoria, British Columbia, V9E 2E7, Canada\\
$^{7}$Department of Astronomy, Shanghai Jiao Tong University, Dongchuan RD 800, 200240 Shanghai, China\\
$^{8}$National Astronomical Observatory of Japan\\
$^{9}$Department of Astronomical Science, The Graduate University for Advanced Studies (Sokendai), 2-21-1, Osawa, Mitaka, Tokyo 181-8588, Japan
$^{10}$Institute for Cosmic Ray Research, University of Tokyo, Japan\\
$^{11}$Department of Astrophysical Sciences, Princeton University,  Princeton, New Jersey, 08544-1001, USA\\
$^{12}$Department of Physics and Astronomy, University of Waterloo, 200 University Avenue West, Waterloo, Ontario, N2L 3G1, Canada \\
$^{13}$Kavli IPMU, University of Tokyo, Japan\\
$^{14}$UCO/Lick Observatory, 1156 High Street, Santa Cruz, California, 95064, USA\\
$^{15}$Department of Physics and Atmospheric Science, Dalhousie University, 6310 Coburg Road, Halifax, Nova Scotia, B3H 4R2, Canada\\
$^{16}$Academia Sinica Institute for Astronomy and Astrophysics, Taipei, Republic of China\\
$^{17}$Research Center for Space and Cosmic Evolution, Ehime University, Bunkyo-cho 2-5, Matsuyama 790-8577, Japan\\
$^{18}$Institute for Advanced Research, Nagoya University, Nagoya 464-8602, Aichi, Japan\\
$^{19}$Department of Astronomy, University of Tokyo, Japan\\
$^{20}$Universit\'e Lyon, Univ Lyon1, Ens de Lyon, CNRS, Centre de Recherche Astrophysique de Lyon UMR 5574, F-69230, Saint-Genis-Laval, France\\
$^{21}$Astronomical Institute, Tohoku University, 2 Chome-1-1 Katahira, Aoba-ku, Sendai, Miyagi-ken, 980-8577, Japan\\
$^{22}$Intsitute for Space and Astronautical Science, Japan Aerospace Exploration Agency, 3-1-1, Yoshinodai, Chuo-ku, Sagamihara, Japan\\
$^{23}$Department of Astronomy \& Astrophysics, University of Toronto, 50 St.\ George Street, Toronto, Ontario, M5S 3H4, Canada \\
\\

\bsp	
\label{lastpage}
\end{document}